\pdfoutput=1
\newcommand*{\ATLASLATEXPATH}{./}
\documentclass[USenglish, texlive=2016, cernpreprint]{\ATLASLATEXPATH atlasdoc}

\newcommand*{\VH}{\ensuremath{VH}\xspace}
\newcommand*{\ttH}{\ensuremath{t\bar{t}H}\xspace}
\newcommand*{\muH}{\ensuremath{\mu_{H}}\xspace}
\newcommand*{\muVBF}{\ensuremath{\mu_{\text{VBF}}}\xspace}

\usepackage[subfigure]{\ATLASLATEXPATH atlaspackage}
\usepackage{\ATLASLATEXPATH atlasbiblatex}

\usepackage{\ATLASLATEXPATH atlascontribute}

\usepackage[process]{\ATLASLATEXPATH atlasphysics}

\graphicspath{{logos/}{figures/}{figures/aux/}}

\usepackage{multirow}
\usepackage{seqsplit}

\AtlasTitle{Search for Higgs bosons produced via vector-boson fusion and decaying into bottom quark pairs in $\sqrt{s} = \SI{13}{\TeV}$ $pp$ collisions with the ATLAS detector}

\author{The ATLAS Collaboration}

\AtlasRefCode{HIGG-2016-30}

\PreprintIdNumber{CERN-EP-2018-140}

\AtlasJournal{PRD}
\AtlasJournalRef{Phys. Rev. D 98 (2018) 052003}
\AtlasDOI{10.1103/PhysRevD.98.052003}

\AtlasAbstract{%
A search for the $b\bar b$ decay of the Standard Model Higgs boson produced through vector-boson fusion is presented.
Three mutually exclusive channels are considered: two all-hadronic channels and a photon-associated channel. 
Results are reported from the analysis of up to \SI{30.6}{\ifb} of $pp$ data at $\sqrt{s} = \SI{13}{\TeV}$ collected with the ATLAS detector at the LHC.  The measured signal strength relative to the Standard Model prediction from the combined analysis is $2.5^{+1.4}_{-1.3}$ for inclusive Higgs boson production and $3.0^{+1.7}_{-1.6}$ for vector-boson fusion production only.
}

\hypersetup{pdftitle={Search for Higgs Bosons Produced via Weak Boson Fusion and Decaying to Bottom Quark Pairs},pdfauthor={The ATLAS Collaboration}}

\begin{document}

\maketitle

\section{Introduction}
\label{sec:intro}
Following the discovery of a new particle with a mass of \SI{125}{\GeV} by the ATLAS and CMS Collaborations at the Large Hadron Collider (LHC)~~\cite{HIGG-2012-27, CMS-HIG-12-028},
there has been an extensive effort to measure its properties and compare them with theoretical predictions for the Standard Model (SM) Higgs boson~\cite{HIGG-2013-01, HIGG-2013-17, HIGG-2013-17, HIGG-2014-11, HIGG-2015-07, CMS-HIG-14-016, CMS-HIG-14-028, CMS-HIG-16-041}.
Precise measurements of the Higgs boson couplings to other SM particles provide insight into the nature of electroweak symmetry breaking since the values of the couplings are determined by the underlying symmetry-breaking mechanism.
The SM Higgs boson production rates and branching ratios are determined by the values of these couplings, and deviations from the predicted values may indicate new particles or forces beyond the Standard Model.
The dominant decay of the SM Higgs boson is into $b\bar b$, but the measurement of Higgs bosons in this decay mode is challenging
because the dominant production mechanisms -- gluon--gluon fusion (ggF) and vector-boson fusion (VBF) -- yield leading-order final states containing only jets.  These hadronic final states are difficult to distinguish from non-resonant $b$-quark production, which has a much larger production rate.
Most previous measurements of $H\rightarrow b\bbar$ decays were made with the relatively rare process of Higgs boson production in association with a leptonically decaying vector boson (\VH, where $V$ denotes a \Wboson or \Zboson boson).
The combined result for a Higgs boson with a mass of \SI{125}{\GeV} from the CDF and D0 experiments is a signal strength $\mu=\sigma/\sigma_\text{SM}=1.9 \pm 0.8$ with a $2.8\sigma$ signal significance~\cite{Aaltonen:2012qt}.
This was followed by measurements with higher significance from ATLAS of $\mu=0.90 \pm 0.27$ at $3.6\sigma$~\cite{HIGG-2016-29} and CMS of $\mu=1.1 \pm 0.3$ at $3.8\sigma$~\cite{CMS-HIG-16-044}.

The VBF process, $pp\rightarrow qqH$, in which the Higgs boson is accompanied by two light-flavor quarks separated by a large rapidity gap,
provides a striking experimental signature for distinguishing Higgs boson production from backgrounds.
A measurement of $H\rightarrow b\bbar$ decay in VBF production mode provides information that is complementary to the measurement in \VH production mode.
The expected production rate $\sigma_\text{VBF} \times B(H\rightarrow b\bar b)$ is \SI{2.2}{\pb}~\cite{Han:1992hr,Djouadi:1997yw,Arnold:2008rz,Bolzoni:2010xr,Bolzoni:2011cu,Cacciari:2024069,Denner:2014cla} at the center-of-mass energy $\sqrt{s} = \SI{13}{\TeV}$. 
Using data collected at $\sqrt{s} = \SI{8}{\TeV}$ and corresponding to an integrated luminosity of \SI{20.2}{\ifb}, the ATLAS experiment set a 95\% confidence level (CL) limit on the production rate of 4.4 times the expected production rate from a VBF-dominated sample with a signal strength $\mu=-0.8 \pm 2.3$ times the predicted value~\cite{HIGG-2014-12}.
The CMS Collaboration used approximately \SI{20}{\ifb} of \SI{8}{\TeV} data to measure a signal strength $\mu = 2.8^{+1.6}_{-1.4}$ corresponding to an observed significance of $2.2\sigma$~\cite{CMS-HIG-14-004}.  

\begin{figure}[htbp]
\centering
\subfigure[]{\includegraphics[width=0.35\textwidth]{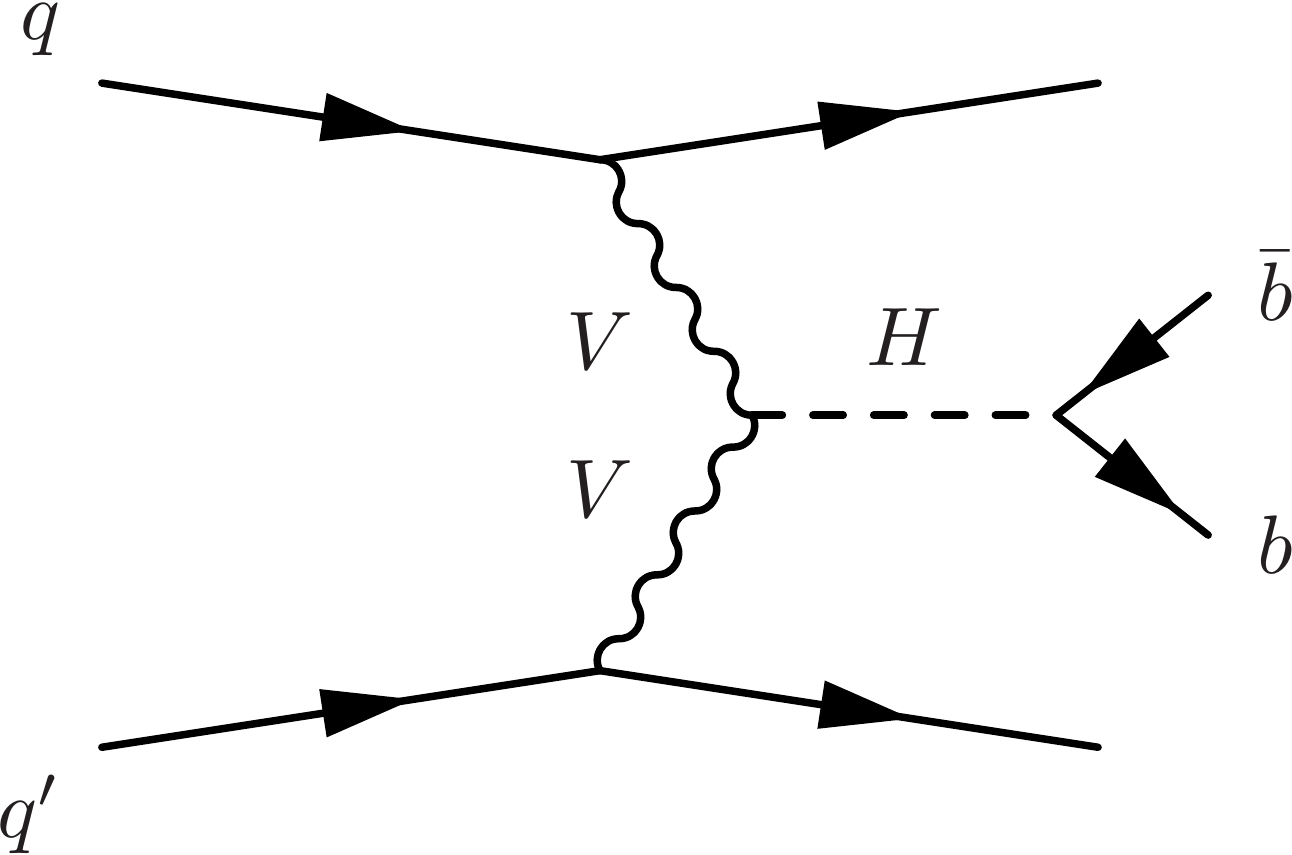}\label{fig:feynman_diagrams_VBF}}
\hspace{0.1 \textwidth}
\subfigure[]{\includegraphics[width=0.35\textwidth]{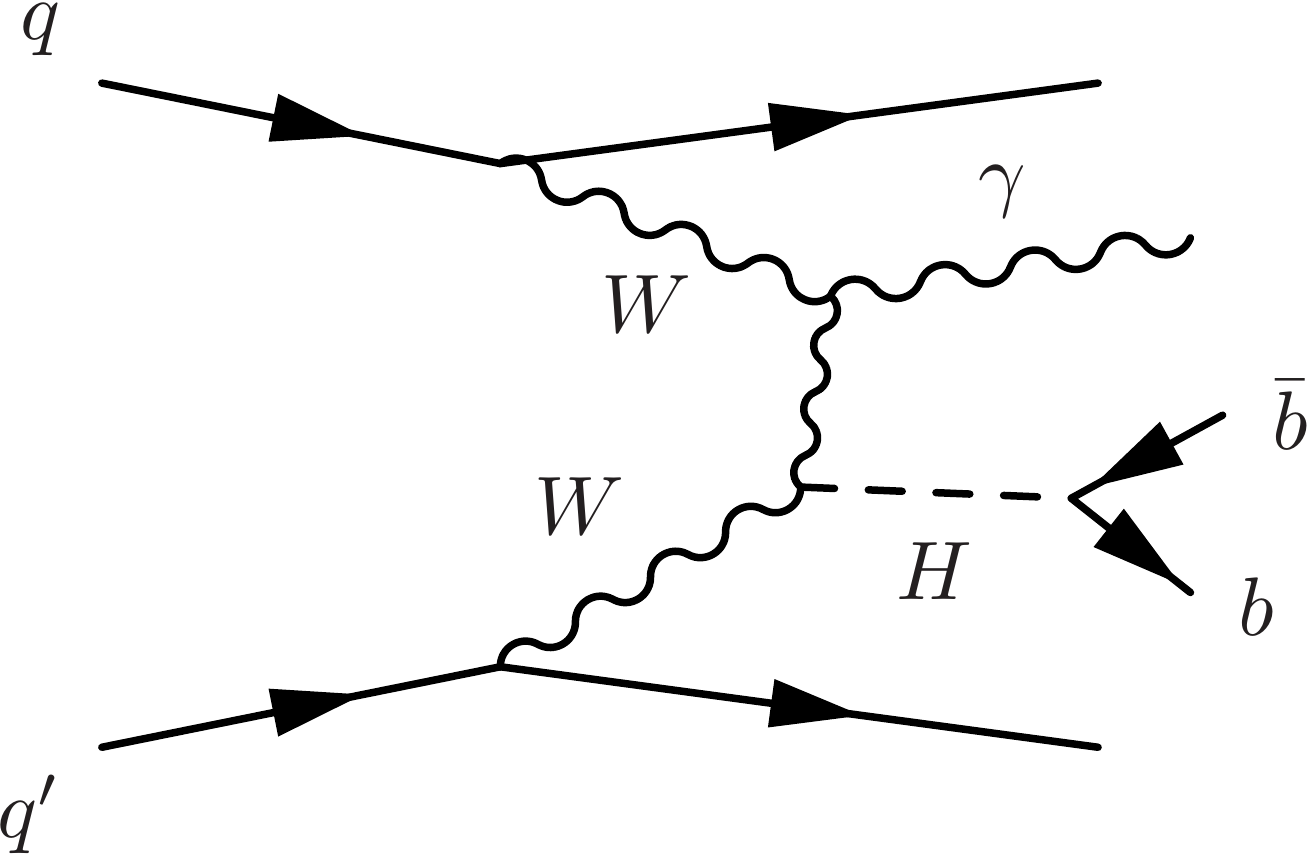}\label{fig:feynman_diagrams_VBFg}}
\caption{Representative Feynman diagrams for Higgs boson production via vector-boson fusion (a) without and (b) with an associated photon. The photon in the right-hand figure may be radiated from any quark leg or charged vector boson.}
\label{fig:feynman_diagrams_signal}
\end{figure}

This article reports the results from a set of complementary search channels sensitive to SM Higgs boson production through VBF with decay into $b\bbar$.
Two of the search channels focus on the process $qqH(\rightarrow b\bbar)$ (Figure~\ref{fig:feynman_diagrams_VBF}) with central and forward jets.
They are collectively referred to as the \textit{all-hadronic} channels because their event selection uses jets only.
The third channel focuses on Higgs boson production in association with a high-momentum photon, $qqH(\rightarrow b\bbar)\gamma$ (Figure~\ref{fig:feynman_diagrams_VBFg}) and is referred to as the \textit{photon} channel.
The presence of an associated photon suppresses the gluon-rich dominant non-resonant $b\bbar jj$ background~\cite{Gabrielli:2007wf, Gabrielli:2016mdd}, further increasing the sensitivity to the VBF final state.
This channel was not included in any previous VBF results.

In all three channels, the signal events are characterized by two central $b$-jets from the decay of a Higgs boson and two light-quark jets with a large rapidity gap between them (VBF jets).
Kinematic properties of the events are used as inputs to boosted decision trees (BDT) trained to classify events as signal-like or background-like.  Backgrounds include contributions from non-resonant jet pairs and from resonant production of $Z$ bosons.  
The signal is extracted from a simultaneous fit to the di-$b$-jet invariant mass ($m_{bb}$) distribution in several regions defined by the BDT discriminant output value. Non-resonant $b\bbar jj$ background and \Zboson (+$\gamma$) + jets background are estimated separately from the fit in each signal region.
Both the inclusive Higgs boson production and the VBF production are measured and presented.
The \textit{all-hadronic} channels are sensitive to ggF as well as VBF, but the photon channel is sensitive to VBF only.
\section{ATLAS detector}
\label{sec:detector}

ATLAS~\cite{PERF-2007-01} is a multipurpose particle detector with a forward--backward symmetric cylindrical geometry and nearly $4\pi$ coverage in solid angle.\footnote{ATLAS uses a right-handed coordinate system with its origin at the nominal interaction point (IP) in the center of the detector and the $z$-axis along the beam pipe. The $x$-axis points from the IP to the center of the LHC ring, and the $y$-axis points upward. Cylindrical coordinates ($r$, $\phi$) are used in the transverse plane, $\phi$ being the azimuthal angle around the beam pipe. The pseudorapidity is defined in terms of the polar angle $\theta$ as $\eta = -\ln \tan(\theta/2)$.}
The interaction point is surrounded by inner tracking devices, a calorimeter system, and a muon spectrometer.

The inner detector provides precision tracking of charged particles for pseudorapidities $|\eta| < 2.5$ and is surrounded by a superconducting solenoid providing a 2~T magnetic field.
The inner detector consists of silicon pixel and microstrip detectors and a transition radiation tracker.
One significant upgrade for the $\sqrt{s} = \SI{13}{\TeV}$ run is the insertable B-layer~\cite{CERN-LHCC-2010-013}, an additional pixel layer close to the interaction point.
It provides high-resolution hits at a small radius to improve tracking performance.

In the pseudorapidity region $|\eta| < 3.2$, high-granularity lead/liquid-argon (LAr) electromagnetic (EM) sampling calorimeters are used to measure EM showers from photons and electrons.
An iron/scintillator tile calorimeter measures hadron energies for $|\eta| < 1.7$.
The endcap and forward regions, spanning $1.5<|\eta| <4.9$, are also instrumented with LAr calorimeters for both the EM and hadronic measurements.

The muon spectrometer consists of a large barrel and two endcap superconducting toroid magnets with eight coils each, a system of trigger chambers, and precision tracking chambers providing triggering and tracking capabilities for muons in the ranges $|\eta| < 2.4$ and $|\eta| < 2.7$, respectively.

A two-level trigger system selects events. The first-level trigger (L1), implemented in hardware, is followed by the software-based high-level trigger, which runs offline reconstruction and calibration software reducing the event rate to less than 1~kHz.

\section{Signal and background simulation}
\label{sec:simulation}
Simulated events are used for signal modeling, BDT training, and background shape determination.
The signal models include both the Higgs boson VBF and ggF production contributions, as well as the small contribution from associated production with top quarks (\ttH) and vector bosons (\VH).
Simulated \textit{all-hadronic} signal events were generated at next-to-leading order in QCD with \POWHEGBOXV{v2}~\cite{Nason:2004rx,Frixione:2007vw,Alioli:2010xd}, using the CT10 parton distribution functions (PDFs)~\cite{Lai:2010vv} and \PYTHIA 8.212~\cite{Sjostrand:2014zea} for parton showering and fragmentation with the AZNLO tuned parameter set~\cite{STDM-2012-23}.
Contributions from \VH and \ttH production were modeled with \PYTHIA 8.212, using the NNPDF PDF~\cite{Ball:2012cx},  and with \MGMCatNLO v2.2.2 ~\cite{Alwall:2014hca} showered with \HERWIGpp 2.7.1~\cite{herwig} and using the NLO CT10 PDF, respectively.
Simulated \Zjets events from strong and electroweak production were generated separately at leading order (LO) plus two partons with \MGMCatNLO v2.2.3 using the NNPDF PDFs and interfaced to \PYTHIA 8.205 with the A14 set of tuned parameters~\cite{ATL-PHYS-PUB-2014-021} for the underlying-event description.
The non-resonant backgrounds in the \textit{all-hadronic} channels are derived exclusively from the data. 

In the \textit{photon} channel, both the $jj\gamma b\bbar$ final-state signal and background events were generated at LO with \MGMCatNLO v2.3.3 using the PDF4LHC\_nlo\_mc PDFs~\cite{Butterworth:2015oua} and interfaced to \PYTHIA 8.212 with the A14 tuned parameter set.  The \VH and \ttH signals were modeled using the same samples as the \textit{all-hadronic} channels.
Background events containing two $b$-quarks from the decay of a \Zboson boson, a photon, and two additional jets were generated separately for strong and electroweak processes.
Non-resonant $\gamma$ + jets simulation events were generated by requiring the same final state as the signal and $Z + \gamma$ background events but excluding diagrams containing on-shell Higgs or \Zboson bosons. The non-resonant $\gamma$ + jets simulation sample is only used in BDT training, while the non-resonant background shape and normalization are obtained from a fit to the $m_{bb}$ data distribution in signal regions (Section~\ref{sec:results}).

Multiple $pp$ collisions were simulated with the soft QCD processes of \PYTHIA 8.186~\cite{Sjostrand:2007gs} using the A2 tuned parameter set~\cite{ATL-PHYS-PUB-2012-003} and the MSTW2008LO PDFs~\cite{Martin:2009iq}.
These additional interactions were overlaid on the hard-scatter interaction of the signal and background samples
according to the luminosity profile of the recorded data to model contributions from $pp$ interactions in both the same bunch crossing and neighboring bunch crossings (pileup).
The response of the ATLAS detector to the generated events was then modeled using full simulation software~\cite{SOFT-2010-01} based on \GEANT4~\cite{Agostinelli:2002hh}, except for the $Z(b\bbar)+\text{jets}$ events, which were passed through a fast simulation where the full calorimeter simulation is replaced by a parameterization of shower shapes~\cite{ATL-PHYS-PUB-2010-013}.

\section{Datasets and object reconstruction}
\label{sec:objects}
This analysis uses LHC $pp$ collision data at a center-of-mass energy of \SI{13}{\TeV} collected 
between September 2015 and October 2016.
The dataset corresponds to an integrated luminosity of  \SI{24.5}{\ifb} for the \textit{all-hadronic} channels and \SI{30.6}{\ifb} for the \textit{photon} channel. The difference in luminosity between the channels is due to limited availability of the triggers for the \textit{all-hadronic} channels during some periods of the data-taking.  The trigger requirements are described in Section~\ref{sec:selection}.  Detector quality requirements are applied to ensure that the selected events are well measured. 
Events are selected using the properties of jets and photons that are reconstructed as described briefly below. 

Jets are reconstructed from topological calorimeter-cell clusters calibrated to the EM scale.
These clusters are inputs to the \antikt jet reconstruction algorithm~\cite{Cacciari:2008gp} with a radius parameter of $R=0.4$.
A likelihood-based discriminant, the Jet Vertex Tagger~\cite{ATLAS-CONF-2014-018}, is applied to jets with transverse momenta $\pT < \SI{60}{\GeV}$ and $|\eta|<2.4$ to suppress jets originating from pileup vertices.
The energy of a jet is corrected using scale factors derived from both the simulated events and an \textit{in situ}, data-based calibration~\cite{PERF-2016-04}
comparing the $\pT$ balance between a jet and a reference object, such as a \Zboson boson, a photon, or a multijet system for various jet-$\pT$ ranges.
In addition, a pileup subtraction algorithm is applied to reduce pileup contributions to the calorimeter-based jet energy.

A flavor-tagging algorithm MV2c10~\cite{PERF-2012-04,ATL-PHYS-PUB-2016-012} tags jets containing $b$-hadrons within the acceptance of the inner detector ($|\eta| < 2.5$)
using log-likelihood ratios from three-dimensional impact parameter significance distributions, secondary vertex information, and the jet \pT and $\eta$.
This information is input to a BDT that calculates the final discriminant.
Three different flavor-tagging operating points are used, corresponding to $b$-tagging efficiencies of 70\%, 77\% and 85\%, respectively,
as measured in simulated $t\tbar$ events for jets having $\pt > \SI{20}{\GeV}$ and $|\eta |< 2.5$~\cite{PERF-2016-05}.
The $c$-jet misidentification efficiencies are measured to be 8.2\%, 16\% and 32\%, respectively, and the light jet misidentification efficiencies are measured to be 0.3\%, 0.7\% and 3.0\%, respectively.
Scale factors are applied to each selected $b$-tagged jet to account for the $b-,c-$ and light--jet flavor-tagging performance differences between data and simulation.
Because the invariant mass $m_{bb}$ is an important discriminant against the non-resonant background,
additional energy corrections are applied to $b$-jets after the jet selection and generic energy calibration. These additional corrections account for semileptonic decays and resolution effects such as energy losses outside of the jet cone~\cite{HIGG-2016-29}.
After these corrections, the full width at half maximum for the signal dijet invariant mass distribution, $m_{bb}$, is \SI{22}{\GeV} for the \textit{all-hadronic} channels and \SI{27}{\GeV} for the \textit{photon} channel.  
The difference is due to the different kinematic requirements for the jets.

Photon reconstruction~\cite{PERF-2013-04} is seeded from clusters of energy deposits in the electromagnetic calorimeter. 
The initial selection based on \textit{loose} criteria uses shower shapes in the second layer of the electromagnetic calorimeter and the energy deposits in the hadronic calorimeter.
The \textit{tight} identification adds information from the finely segmented first layer of the calorimeter, which provides good rejection of hadronic jets in which a neutral meson carries most of the jet energy.
Clusters without any matching track or conversion vertex are classified as unconverted photon candidates.  Clusters with a matching vertex reconstructed from one or two tracks are converted photon candidates.
Both the converted and unconverted photon candidates with transverse energy $\ET > \SI{30}{\GeV}$ in the pseudorapidity ranges $|\eta|<1.37$ or $1.52 < |\eta| < 2.37$ are used. The range $1.37 < |\eta| < 1.52$  is excluded because it is the gap between the barrel and endcap sections of the calorimeter.  
To further suppress hadronic background from jets and neutral pions, an isolation requirement is applied to the photon candidates.
The calorimeter isolation variable $\ET^\text{iso}$ is the sum of the transverse energy of three-dimensional positive-energy topological clusters~\cite{PERF-2014-07} reconstructed in the electromagnetic and hadronic calorimeters in a cone of size $\Delta R \equiv \sqrt{(\Delta\phi)^2+ (\Delta\eta)^2} =0.4$ around the photon candidate,
where the $\Delta\eta\times\Delta\phi$ region of size $0.125\times 0.175$ around the photon cluster's centroid is excluded.
The isolation requirement, which depends explicitly on the photon transverse energy $\ET^\gamma$, is $\ET^\text{iso} < \SI{2.45}{\GeV} + 0.022 \times \ET^\gamma$.
This requirement provides a signal efficiency around 98\% over the \ET range expected for the \textit{photon} channel.

Electrons are reconstructed~\cite{ATLAS-CONF-2016-024} with a sliding-window algorithm based on the clusters of energy deposits in the electromagnetic calorimeter and matched to the tracks from the inner detector.
Electron candidates must satisfy the \textit{tight} likelihood-based electron identification criteria~\cite{PERF-2016-01}, which combine the requirements of calorimeter shower shape, track-to-cluster association, and associated track qualities.
Identified electrons are required to pass track- and calorimeter-based isolation requirements and to have $\ET > \SI{27}{\GeV}$ and $|\eta|<2.47$.
The track-based isolation requirement is a function of the electron \pT and is based on the other tracks around the electron-associated track within a variable cone size up to $\Delta R=0.2$.
The calorimeter-based isolation criterion requires the sum of transverse energies of clusters not associated with an electron candidate within a cone of $\Delta R=0.2$ around the electron track to be smaller than \SI{3.5}{\GeV}.

Muons are reconstructed~\cite{PERF-2015-10} by combining the inner detector and muon spectrometer measurements up to $|\eta|=2.5$.
Muon candidates are required to have $\pT > \SI{25}{\GeV}$ and satisfy the \textit{medium} muon identification criteria~\cite{PERF-2015-10}.
Identified muons must pass an isolation selection requiring the sum of transverse momenta of tracks within a cone of $\Delta R=0.2$ around the muon track, excluding the muon candidate, to be smaller than \SI{1.25}{\GeV}.

Double-counting of photons, leptons, and jets is avoided by applying an overlap removal algorithm based on the $\Delta R $ distance metric.
First, jets within $\Delta R = 0.2$ of any identified photons, muons or electrons are removed.
Then, any photons, muons and electrons that lie $0.2 < \Delta R < 0.4$ from the jet axis are removed.
Finally, photons within $\Delta R = 0.4$ of an identified muon are removed, and electrons within $\Delta R = 0.4$ of an identified photon are removed.

\section{Event selection}
\label{sec:selection}
The event selection targets three distinct final-state topologies: two \textit{all-hadronic} channels and the \textit{photon} channel.
The selection criteria are matched to a set of dedicated trigger algorithms used to identify events compatible with VBF $H\ra b\bbar$ production.
In the following, the central region corresponds to $|\eta|<2.8$, and the forward regions correspond to the range $3.2 < |\eta|<4.4$.
The channel definitions are
\begin{itemize}
   \item \textit{two-central}: at least one VBF jet is required to be in the forward region, and both $b$-tagged jets from the Higgs boson decay are in the central region,
\item \textit{four-central}: both VBF jets and both $b$-tagged jets from the Higgs boson decay are found in the central region of the detector, and
\item \textit{photon}: a photon and both $b$-tagged jets from the Higgs boson decay are found in the central region and both VBF jets are within the detector acceptance.
\end{itemize}
The selected events for the two \textit{all-hadronic} channels are mutually exclusive. 
The small overlap between the \textit{photon} and \textit{all-hadronic} channels is removed with an explicit veto of any data events in the \textit{all-hadronic} selection passing the \textit{photon} selection.
The 0.5\% overlap in the simulated signal sample is ignored.

\subsection{\textit{Two-central} channel trigger and event selection}

The \textit{two-central} channel requires a central jet with $\ET > \SI{40}{\GeV}$, another central jet with $\ET > \SI{25}{\GeV}$, and a forward jet with $\ET > \SI{20}{\GeV}$ to pass the L1 trigger.

In the high-level trigger, one central $b$-tagged jet~\cite{TRIG-2016-01} at the 70\% $b$-tagging efficiency working point with $\ET > \SI{80}{\GeV}$,
another central $b$-tagged jet at the 85\% $b$-tagging efficiency working point with $\ET > \SI{60}{\GeV}$, and a forward jet with \ET at least \SI{45}{\GeV} are required.  The same $b$-tagging algorithm is used in the online selection as in the offline selection.  

Selected events must have at least four offline reconstructed jets with $\pT > \SI{20}{\GeV}$ and $|\eta|<4.4$.
Among the selected jets, at least one jet must have $\pT > \SI{95}{\GeV}$, have $|\eta|<2.4$, and pass the 70\% $b$-tagging efficiency working point requirement. 
The $|\eta|$ requirement is narrower than the nominal requirement for $b$-tagging ($|\eta|<2.5$) for comparison with a supporting trigger used for validation.   
At least one additional jet is required to pass the 85\% $b$-tagging efficiency working point selection and have $\pT > \SI{70}{\GeV}$ and $|\eta|<2.5$.
Finally, events are required to have at least one forward jet with $\pT > \SI{60}{\GeV}$.
These thresholds were determined by the efficiency plateau of the trigger-jet transverse-energy requirements.
The two highest-\pT $b$-tagged jets are chosen to form the Higgs boson candidate.
Among the remaining jets, the two jets with highest invariant mass including at least one forward jet are designated as the VBF jets.

\subsection{\textit{Four-central} channel trigger and event selection}

The \textit{four-central} channel requires four central jets to pass the L1 trigger with $\ET > \SI{15}{\GeV}$.

The requirements of the high-level trigger varied during the course of data-taking.  In the first half, events were required to have two $b$-tagged jets with $\ET>\SI{45}{\GeV}$~passing the 70\% efficiency working point requirements for the trigger $b$-tagging algorithm. In the second half, the trigger's jet $\ET$ thresholds were changed to \SI{35}{\GeV} and the $b$-tagging algorithm was tightened to operate at 60\% efficiency to achieve an overall lower rate of events passing the high-level trigger.

The selected events are required to have at least four jets reconstructed with offline algorithms with $\pT > \SI{55}{\GeV}$ and $|\eta|<2.8$ to match the trigger requirements.
At least two jets must pass the 70\% $b$-tagging efficiency working point requirement. 
All $b$-tagged jets must be within the acceptance of the inner detector ($|\eta| < $ 2.5).  The two highest-\pT $b$-tagged jets form the Higgs boson candidate.
Among the remaining jets, the pair of non-$b$-tagged jets with highest invariant mass is taken as the VBF jet pair.
Finally, events containing at least one forward jet with $\pT > 60~\GeV$ are removed to avoid overlap with the \textit{two-central} channel.

\subsection{\textit{Photon} channel trigger and event selection}

The \textit{photon} channel requires a photon to pass the L1 trigger with $\ET > \SI{22}{\GeV}$.
In the high-level trigger, a photon with $\ET > \SI{25}{\GeV}$ is required in addition to
 at least four jets with $\ET > \SI{35}{\GeV}$ and $|\eta|<4.9$, and at least one dijet pair with invariant mass greater than \SI{700}{\GeV}.
For the first half of the data-taking, no online $b$-tagging requirements were applied;
for the second half, which had increased instantaneous luminosity, at least one jet is required to be $b$-tagged at the 77\% efficiency working point.

The event selection for the \textit{photon} channel requires a photon with $\ET > \SI{30}{\GeV}$ in the calorimeter regions $|\eta|<1.37$ or $1.52<|\eta|<2.37$.
Events must have at least four jets, all satisfying $\pT > \SI{40}{\GeV}$ and $|\eta|<4.4$, with at least two jets in $|\eta|<2.5$ passing the 77\% $b$-tagging efficiency working point requirement. 
The two highest-\pT $b$-tagged jets are taken to be the signal jets of the Higgs boson decay.
Among the remaining jets, the pair with the highest invariant mass is chosen to be the VBF jet pair.
The invariant mass of the VBF jets is required to be at least \SI{800}{\GeV} so that the trigger requirement imposed on the invariant mass is fully efficient.

\subsection{ The $bb$-system \pT requirement}

The jet \pT thresholds in the trigger and the offline selection sculpt the $m_{bb}$ distribution.
To remove this sculpting, which could bias the final $m_{bb}$ fit, the $bb$-system $\pt$ in the \textit{two-central},  \textit{four-central}, and \textit{photon} channels is required to be larger than 150, 160, and \SI{80}{\GeV}, respectively.

The full event selection, including the trigger requirements and offline requirements, is summarized in Table~\ref{tab:evtsel}.

\begin{table}[htbp]
   \caption{Trigger and event selection criteria for all search channels. L1 and HLT refer to the first-level trigger and the high-level trigger, respectively. The \pT~ and $|\eta|$ requirements on the offline jets are used to match trigger selections and flavor-tagging requirements.
   All the selection criteria are applied independently.}
   \label{tab:evtsel}
   \def\arraystretch{1.4}
   \centering
      {\footnotesize
         \begin{tabular}{ l | l | l }
            \hline \hline
            \multicolumn{3}{c}{ \textit{Two-central} channel} \\
            \hline
        \multirow{4}{*}{Trigger}    &\multirow{2}{*}{L1}    & $\geq2$ central jets with $\ET > 40, 25$ \GeV{} \\
                                    &                         & $\geq1$ forward jet with $\ET > 20$ \GeV{} \\
            \cline{2-3}
                                    &\multirow{2}{*}{HLT}   & $\geq2$ central $b$-jets at 70\%, 85\% efficiency working points with $\ET > 80, 60$ \GeV{} \\
                                    &                         & $\geq1$ forward jet with $\ET > 45$ \GeV{} \\
            \hline
            \multicolumn{2}{c|}{\multirow{4}{*}{Offline}} & $\geq2$ $b$-jets at 70\%, 85\% efficiency working points with $\pt>95, 70$ \GeV{} and $|\eta|<2.5$  \\
            \multicolumn{2}{c|}{}                         & $\geq1$ jet with $\pt>60$ \GeV{} and $3.2<|\eta|<4.4$  \\
            \multicolumn{2}{c|}{}                        & $\geq1$ jet with $\pt>20$ \GeV{} and $|\eta|<4.4$ \\
            \multicolumn{2}{c|}{}                       &  $\pt(bb)>160$ \GeV{} \\
            \hline

            \multicolumn{3}{c}{\textit{Four-central} channel} \\
            \hline
            \multirow{2}{*}{Trigger} & L1                    & $\geq4$ central jets with $\ET > 15$ \GeV{} \\
            \cline{2-3}
			                         & HLT   & $\geq2$ central $b$-jets at 70\% (or 60\%) efficiency working point with $\ET > 45$ \GeV{} (or 35 \GeV{}) \\
            \hline
            \multicolumn{2}{c|}{\multirow{4}{*}{Offline}} & $\geq2$ $b$-jets at 70\% efficiency working point with $\pt>55$ \GeV{} and $|\eta|<2.5$\\
            \multicolumn{2}{c|}{}                         & $\geq2$ jets with $\pt>55$ \GeV{} and $|\eta|<2.8$  \\
            \multicolumn{2}{c|}{}                         & No jet with $\pt>60$ \GeV{} and $3.2<|\eta|<4.4$    \\
            \multicolumn{2}{c|}{}                        & $\pt(bb)>150$ \GeV{}                       \\
            \hline

            \multicolumn{3}{c}{ \textit{Photon} channel} \\
            \hline
       \multirow{4}{*}{Trigger} &     L1                    & $\geq1$ photon with $\ET > 22$ \GeV{} \\
            \cline{2-3}
                                & \multirow{3}{*}{HLT}  & $\geq1$ photon with $\ET > 25$ \GeV{}\\
								&         & $\geq4$ jets (or $\geq3$ jets and $\geq1$ $b$-jet at 77\% efficiency working point) with $\ET>35$ \GeV{} and $|\eta|<4.9$  \\
                                &                         & $m_{jj}>$ 700 \GeV{} \\
            \hline
           \multicolumn{2}{c|}{\multirow{5}{*}{Offline}} & $\geq 1$ photon with $\ET > 30$ \GeV{} and $|\eta|<1.37$ or $1.52 < |\eta|<2.37$\\
           \multicolumn{2}{c|}{}                         & $\geq2$ $b$-jets at 77\% efficiency working point with $\pt>40$ \GeV{} and $|\eta|<2.5$  \\
           \multicolumn{2}{c|}{}                         & $\geq2$ jets with $\pt>40$ \GeV{} and $|\eta|<4.4$  \\
           \multicolumn{2}{c|}{}                         & $m_{jj} >$ 800 \GeV{}      \\
           \multicolumn{2}{c|}{}                        & $\pt(bb)>80$ \GeV{}                         \\

            \hline \hline
         \end{tabular}
      }
\end{table}

\section{Multivariate analysis}
\label{sec:mva}
After the event selection requirements are applied, a set of BDTs classify events as being signal-like or background-like~\cite{scikit-learn,tmva}. 
A separate BDT is trained for each channel with the AdaBoost~\cite{FREUND1997119} algorithm.
Each BDT discriminant is constructed from a set of variables to maximize the separation between the signal and the dominant backgrounds. 
The discriminant is then used to define event categories of varying signal purity.

Since the observed signal is extracted from the $m_{bb}$ spectrum, the input variables for each BDT are chosen to have low correlation with the $bb$ invariant mass to prevent sculpting of the distribution.
The number of input variables is minimized by excluding variables that give only marginal performance improvement.
The following input variables are used for all channels, with $j1$ and $j2$ denoting the leading and sub-leading \pT\ VBF jets and with $b1$ and $b2$ denoting leading and sub-leading \pT\ Higgs boson $b$-jet candidates:
\begin{itemize}
\item $m_{jj}$: the invariant mass of the VBF jet pair.
\item $\pT^{jj}$: the transverse momentum of the VBF jet pair.
\item $N_\text{trk}^{j1}$, $N_\text{trk}^{j2}$: the number of tracks with $\pT > \SI{0.5}{\GeV}$ in the VBF jets, $j1$ and $j2$.  This variable discriminates between gluon jets, which are more abundant in the background processes, and light-quark jets, which are present in the signal. The variable is only used for jets with $|\eta| <$ 2.5.
\item $\pT^\text{balance}$: the ratio of the vectorial and scalar sums of the jet (and photon, if applicable) transverse momenta,
$$\frac{\left| \overrightarrow{\pT}(j1)+\overrightarrow{\pT}(j2)+\overrightarrow{\pT}(b1)+\overrightarrow{\pT}(b2) \right| }{ \pT (j1)+\pT(j2)+\pT(b1)+\pT(b2)  }.$$
This variable discriminates between electroweak signal processes, which typically are balanced, and multijet QCD events, which are less balanced.
\item $\cos{\theta}$: cosine of the angle between the normal directions of the planes spanned by the VBF jet pair and signal $b$-jet pair in the center-of-mass frame of the $jjbb$ system, which is related to the angular dynamics of the production mechanism.
\end{itemize}

In addition to these common variables, the \textit{two-central} and \textit{four-central} channel BDTs include the following input variables:
\begin{itemize}
\item $\text{max} (\eta) \equiv \max( |\eta_{j1}|, |\eta_{j2}| )$: the maximum absolute value of the VBF jet pseudorapidity.
\item $\eta^* = \frac{1}{2}( |\eta_{j1}| + |\eta_{j2}| - |\eta_{b1}| - |\eta_{b2}|)$: the average pseudorapidity difference between VBF and signal jets.  This variable discriminates between QCD multijet events, which have no average pseudorapidity difference, and VBF processes, where the VBF jets are on average more forward than the signal jets.    
\item $\min\Delta R(j1)$: minimum angular separation between the leading VBF jet and the closest jet with $\pT > \SI{20}{\GeV}$ and $|\eta| <4.4$ which is not a signal or VBF jet.
\item $\min\Delta R(j2)$: minimum angular separation between the sub-leading VBF jet and  the closest jet with $\pT > \SI{20}{\GeV}$ and $|\eta| <4.4$ which is not a signal or VBF jet.
\item $\Delta m_{jj}$: the difference between the invariant mass of the VBF jet pair and the largest invariant mass of any jet pair in the event, excluding the two jets forming the Higgs boson candidate.
\end{itemize}

The \textit{photon} channel BDT includes the following variables in addition to the common variables:
\begin{itemize}
   \item $\Delta R(b1,\gamma)$, $\Delta R(b2,\gamma)$: angular separation between the signal $b$-jets and the photon.
   \item $\Delta \eta_{jj}$: $\eta$ separation between the VBF jets.
   \item $\text{centrality}(\gamma,jj)$: centrality of the photon relative to the VBF jets:
   $$\text{centrality}(\gamma,jj)=\left| \frac{y_{\gamma}-\frac{y_{j1}+y_{j2}}{2}}{y_{j1}-y_{j2}} \right| , $$
    where $y$ is the rapidity.
\item $\Delta\phi(bb,jj)$ : azimuthal angle between the VBF jet pair and the signal $b$-jet pair.
\end{itemize}

The training signal samples are the VBF signal simulation samples described in Section~\ref{sec:simulation}.
For the \textit{two-central} and \textit{four-central} channel BDTs, the training background sample is a set of data events in the mass sidebands $\SI{80}{\GeV} < m_{bb} < \SI{100}{\GeV}$ and $\SI{150}{\GeV}< m_{bb} < \SI{190}{\GeV}$.  
Events from \Zjets production are not removed; however, they contribute 2(3)\% to the low mass sideband for the \textit{two(four)-central} channel.  
Because data from the sidebands are used as the training sample for the hadronic channels, a three-fold validation of the BDT training is performed to verify possible over-training.
The signal and background samples are randomly divided into three equal subsets which are then each used to train the BDT while the other two subsets are used for testing. 
Equal discriminating power is found across all subsets.   The effect of potential bias on the $m_{bb}$ distribution  from over-training  was checked by repeating the analysis in the training and validation sets.  Additionally, Asimov datasets were produced by reweighting the sidebands to the observed difference in the ratio of events in the training and validation samples for each region.  Observed biases in these tests were negligible compared to the total statistical error on the signal.
In the \textit{photon} channel, there are not enough data events to form a training sample, so the non-resonant $\gamma$ + jets simulation sample is used as the background training sample. 
The entire set of events is split into two samples for training and evaluation.
To validate the modeling of the non-resonant QCD background, the simulated events are compared with the data events in the mass sidebands ($m_{bb}< \SI{100}{\GeV}$ and $m_{bb}>\SI{140}{\GeV}$). 
Correction functions are applied to reweight some of the kinematic distributions ($\Delta \eta_{jj}$, $\pT^{jj}$, $\pT^{\text{balance}}$, and minimum of $\Delta R(b1,\gamma)$ and $\Delta R(b2,\gamma)$) to improve the overall modeling by the non-resonant QCD simulation sample.
The reweighting process is performed iteratively. 
The correction function is determined for the kinematic distribution of one of the input variables.
This function is then applied back to the non-resonant $\gamma$ + jets simulation sample to reweight the kinematic distribution of the input variable
that the correction function is determined for. 
At the same time, the kinematic distributions of other input variables correlated with this input variable
also show improved agreement with the data after the reweighting.
The distributions of the other uncorrelated input variables are not affected by the reweighting.
The process is repeated to reweight the other three kinematic variables.

\begin{figure}[htbp]
\centering
\includegraphics[width=0.48\textwidth]{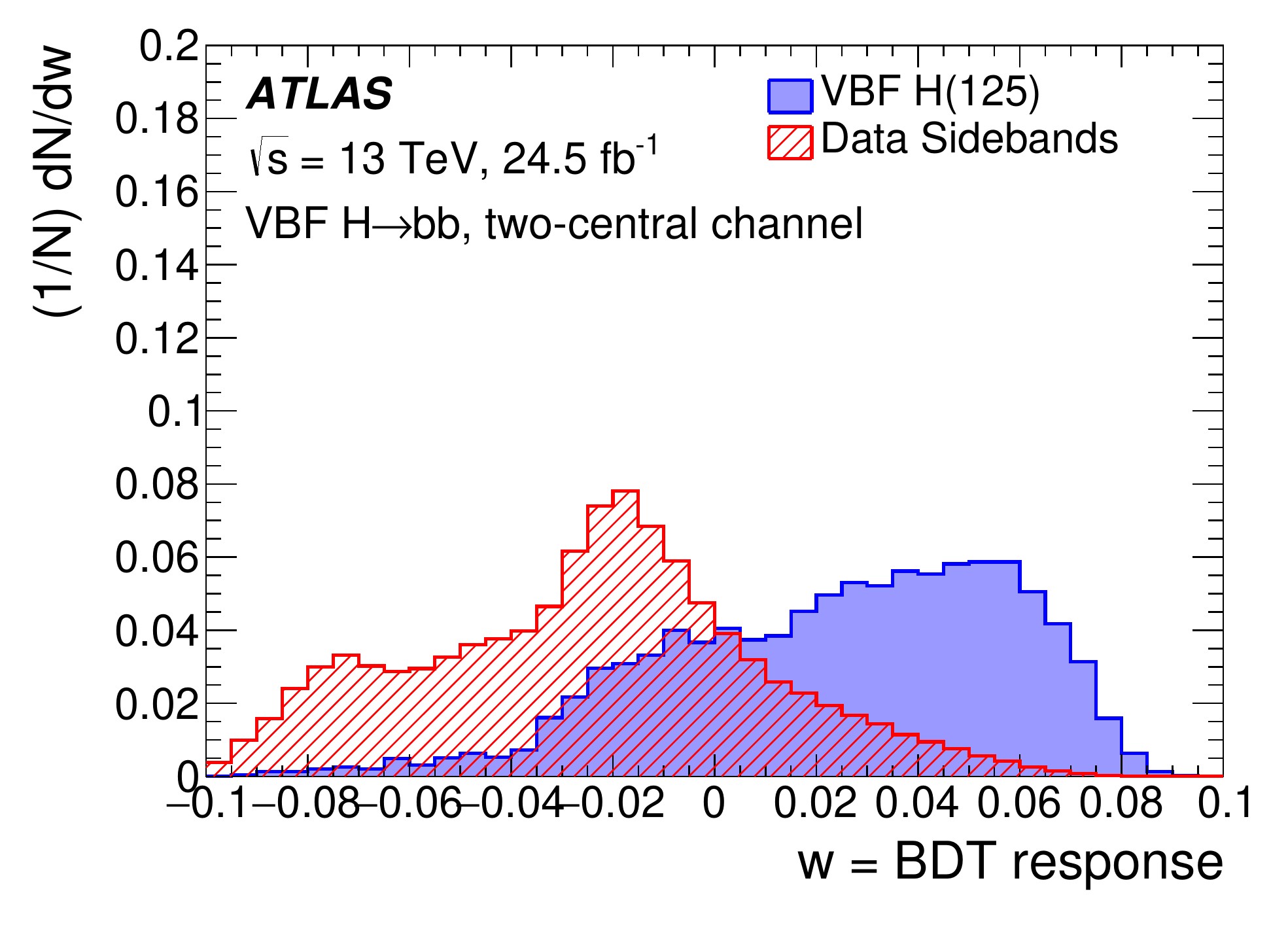}
\includegraphics[width=0.48\textwidth]{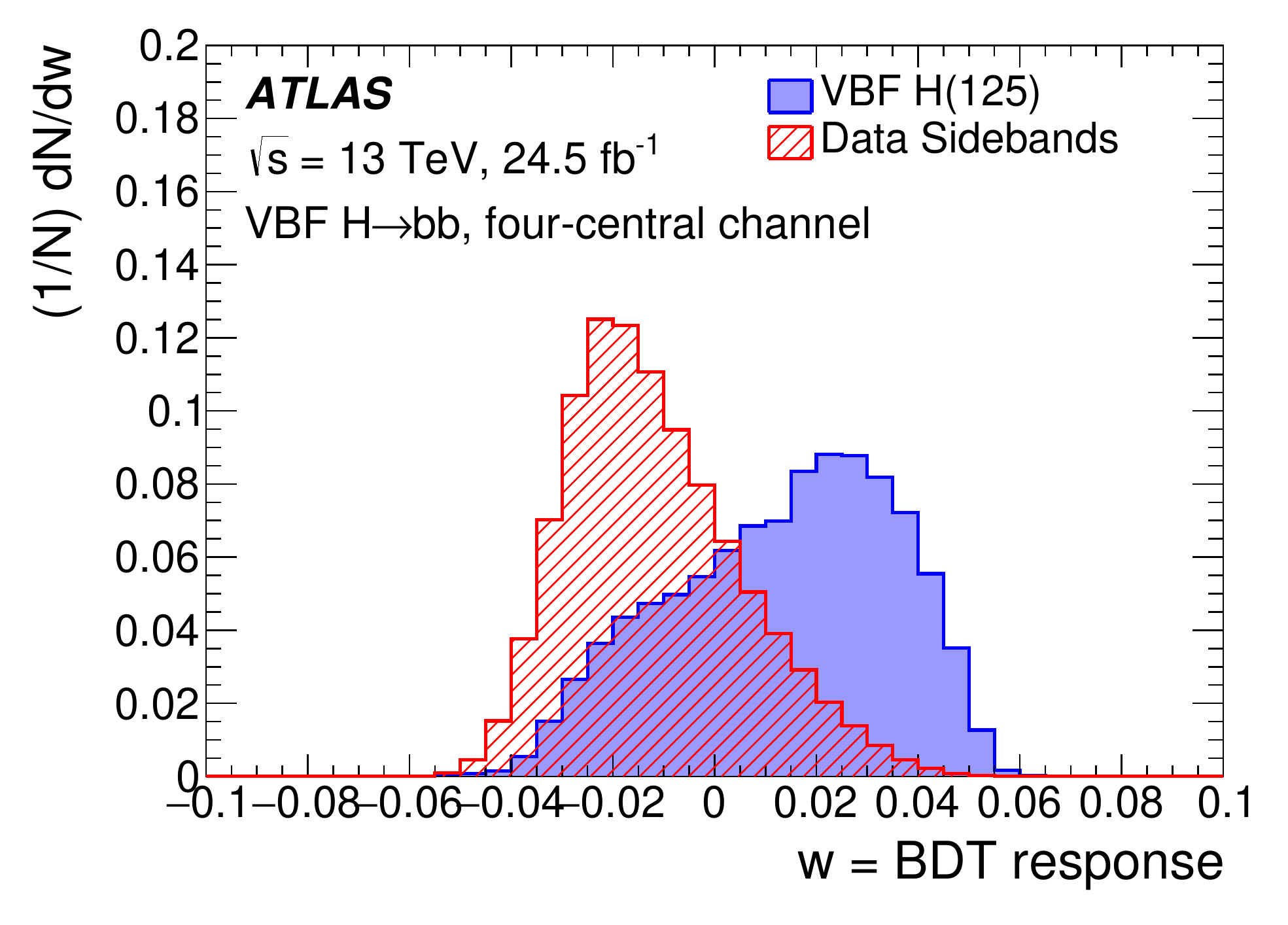}
\includegraphics[width=0.48\textwidth]{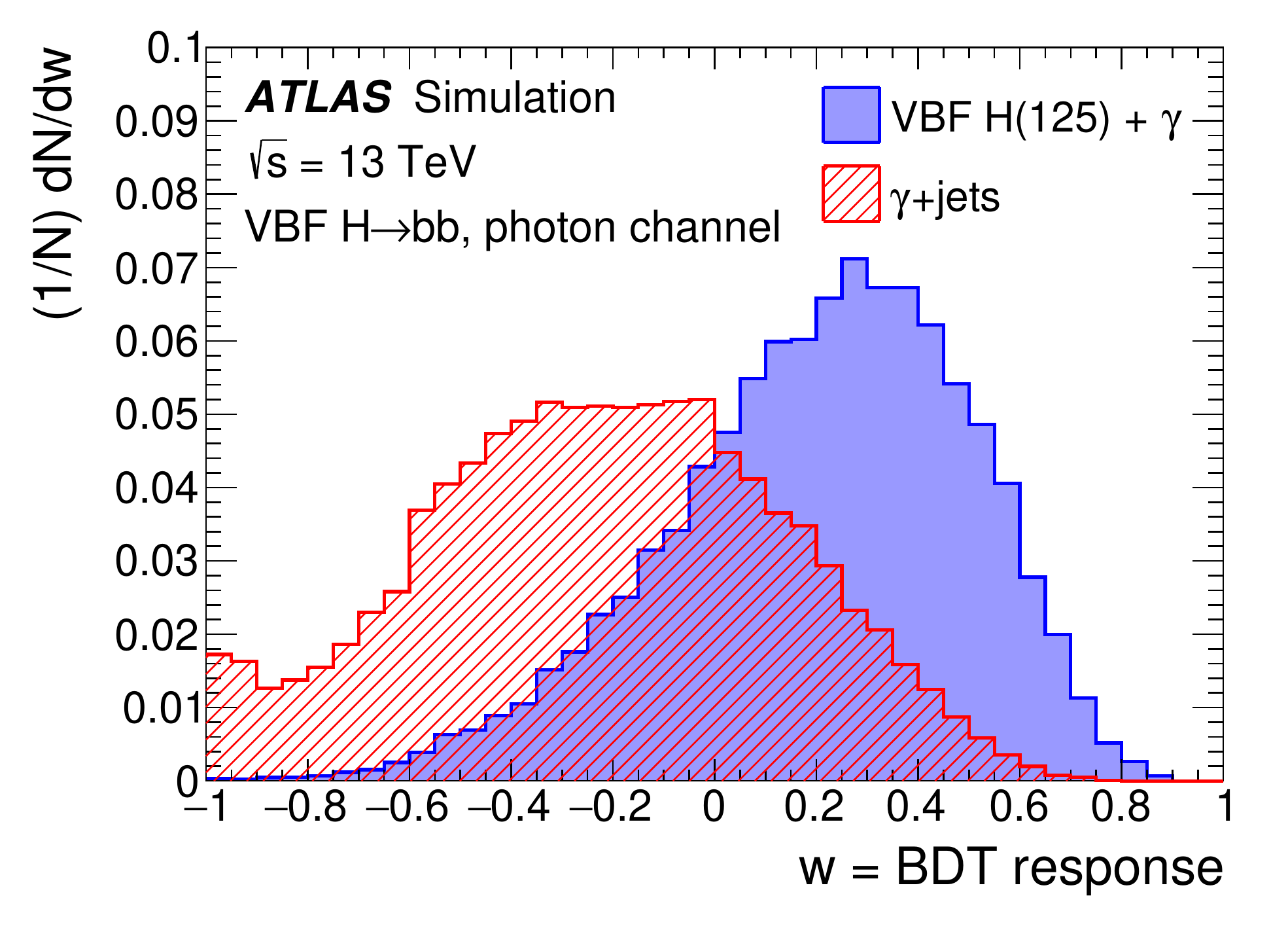}
\caption{The BDT response $w$ for the signal and dominant backgrounds, normalized to unity.  The \textit{two-central} channel is on the top left, the \textit{four-central} channel is on the top right, and the \textit{photon} channel is shown in the bottom row.
For these plots, the background refers to the continuum background derived either from the $m_{bb}$ sidebands in data (\textit{all-hadronic} channels) or from the $\gamma$ + jets simulation sample (\textit{photon} channel).  }
\label{fig:bdt_output_comparison}
\end{figure}

The BDT responses for the signal and background samples are shown in Figure~\ref{fig:bdt_output_comparison}. 
The output discriminant from each BDT is used to define several signal regions (SR).  The \textit{two-central} channel has two regions, the \textit{four-central} has four regions, and the \textit{photon} channel has three regions as summarized in Table~\ref{tab:region_definition}.  The \textit{four-central} channel does not include the full BDT range in the set of signal regions because including events with lower BDT response did not improve the significance of the result.  
These region definitions are optimized for sensitivity to the Higgs boson signal while limiting the maximum experimental uncertainty of the \Zboson boson contribution in any signal region to less than 1.5 times the Standard Model \Zboson boson prediction. The requirement on the precision of the \Zboson boson contribution is necessary because this contribution is a parameter in the fit and the signal regions must be large enough to ensure it is well measured.

\begin{table}[hbtp]
\caption{Criteria for the BDT responses used to define the signal regions (SR) for the three channels.}
\label{tab:region_definition}
\centering
\begin{tabular}{l|cccl}
\hline
 \hline
 Region                & SR IV           & SR III         & SR II          & \multicolumn{1}{c}{SR I}   \\
\hline
\textit{Four-central} & (0.002, 0.015] & (0.015, 0.026] & (0.026, 0.033] & $> 0.033$       \\
\textit{Two-central}  &                &                & $<-0.006$    & $\geq-0.006$    \\
\textit{Photon}       &                & $<-0.05$       & [$-0.05$, 0.30] & $>0.30$         \\
 \hline
\hline
\end{tabular}
\end{table}

\section{Background and signal modeling}
\label{sec:backgrounds}
The main sources of background contributing to the final-state signatures are divided into two groups: processes with decay of a massive particle into $b$-tagged jet pairs and processes with non-resonant $b$-tagged jet pairs.
The resonant backgrounds are dominated by \Zboson (+$\gamma$) + jets, with small contributions from \Wboson (+$\gamma$) + jets.
The non-resonant backgrounds are dominated by multijet (+$\gamma$) production, with small contributions from $\ttbar$ (+$\gamma$) and single-top events.
For all of the backgrounds, $b$-tagged jets may correspond to true $b$-jets or to misidentified $c$-jets, $\tau$-jets, or light-flavor jets.
Both the background and signal $m_{bb}$ shapes are parameterized with functions that are derived differently depending on whether they arise from a resonant or non-resonant process. The contributions from \Zboson (+$\gamma$) + jets and multijet (+$\gamma$) production background processes are derived from fits to the $m_{bb}$ data distribution using template distributions or analytical functions constructed from sideband data regions or simulation samples. The contributions from other background processes are estimated from simulations.

The Higgs boson and $Z\rightarrow \bbbar$ resonance shapes are parameterized with histogrammed Bukin functions~\cite{bukin} (\textit{two-central}  and \textit{four-central} channels) or  Crystal Ball functions~\cite{cb1,cb2} (\textit{photon} channel).
In general, the $m_{bb}$ distributions are well modeled by these functions, and a closure test performed on a representative test dataset, called an Asimov dataset~\cite{Cowan:2010js}, composed of these distributions plus the non-resonant background indicates no bias in the extracted signal normalization.

In all channels, the non-resonant background distributions are modeled as polynomials and derived from data.  
In the \textit{two-central} and \textit{four-central} channels, Bernstein polynomials are fit to the sidebands of the $m_{bb}$ distribution outside the signal region of $\SI{100}{\GeV} < m_{bb} < \SI{140}{\GeV}$ for each BDT region separately.   
The \textit{photon} channel uses a general polynomial.  
The \Zjets contribution is subtracted using predictions from simulations; tests showed that subtracting twice the prediction does not change the chosen function. 
For all channels, the lowest-order polynomial which satisfies basic goodness-of-fit requirements, including $\chi^2$ and $F$-tests, is chosen as a candidate.
Using Asimov datasets derived from alternative background parameterizations which also satisfy these criteria in fits to the data sidebands, an additional function-selection criterion is applied to ensure that the chosen function candidate does not create any significant spurious signal.  
This criterion is that any function which induces a spurious Higgs boson signal contribution with an absolute signal strength of one or larger in these Asimov datasets is discarded. 
This requirement minimizes the possibility that the chosen function could generate a signal.  
The alternative functions used include a product of Bernstein polynomials and exponential functions, as well as a sum of exponential functions. 
A third-order Bernstein polynomial is the lowest-order polynomial which satisfies these criteria for the \textit{two-central} and \textit{four-central} channel regions, except SR IV of the \textit{four-central} channel, which requires a fourth-order Bernstein polynomial. 
The \textit{photon} channel uses a second-order polynomial.

The \Zboson($\rightarrow \bbbar$) + jets contribution plays an important role in the fit procedure because it contributes to the low $m_{bb}$ sideband and affects the continuum background determination.
Studies have shown that typical methods of estimating uncertainties for a leading-order \Zboson + jets simulation may not be appropriate in the regions of phase space used in this analysis, including high-\pT~boson production with widely separated jets~\cite{Rubin:2010xp}.  
Therefore, its normalization is allowed to float independently in each BDT region.   
In the \textit{two-central} and \textit{four-central} channels, the low mass sideband extends only to \SI{80}{\GeV} due to the trigger thresholds.   
Therefore, it does not provide a strong constraint on the \Zboson($\rightarrow \bbbar$) + jets contribution, and consequently the determination of this background contributes significantly to the overall uncertainty.  
In the \textit{photon} channel, the sidebands extend to \SI{50}{\GeV}, allowing the fit to provide a strong constraint on the \Zboson($ \rightarrow \bbbar$) + jets contribution.

A summary of the estimated number of signal events in the Higgs boson mass window of $\SI{100}{\GeV} < m_{bb} < \SI{140}{\GeV}$ is given in Table~\ref{tab:expected_signal}.  There is up to a 60\% contribution of ggF events to the Higgs boson signal in the  \textit{all-hadronic} channels in the least sensitive signal regions and up to a 20\% contribution in the \textit{photon} channel.

\begin{table}[htbp]
\caption{Expected numbers of signal events within the Higgs boson mass window of $\SI{100}{\GeV} < m_{bb} < \SI{140}{\GeV}$ estimated from simulations. Statistical uncertainties are shown for the predictions from simulations.}
\label{tab:expected_signal}
\centering
\resizebox{\textwidth}{!}{
   \begin{tabular}{ l |r@{$\pm$}lr@{$\pm$}l|r@{$\pm$}lr@{$\pm$}lr@{$\pm$}lr@{$\pm$}l|r@{$\pm$}lr@{$\pm$}lr@{$\pm$}l}
\hline
\hline
Channel      & \multicolumn{4}{c|}{\textit{two-central}} & \multicolumn{8}{c|}{\textit{four-central}} & \multicolumn{6}{c}{\textit{photon}} \\ \hline
Region       & \multicolumn{2}{c|}{SR I}           & \multicolumn{2}{c|}{SR II  }         & \multicolumn{2}{c|}{SR I }  & \multicolumn{2}{c|}{SR II} & \multicolumn{2}{c|}{SR III} & \multicolumn{2}{c|}{SR IV}  & \multicolumn{2}{c|}{SR I}  & \multicolumn{2}{c|}{SR II}   & \multicolumn{2}{c}{SR III}   \\
\hline
 VBF   & 101.2&2.0 & 22.2 &0.9 & 51.6 &1.1 & 28.4 &0.9 & 43.1&1.0 & 41.9 &1.1 & 6.2 &0.1 & 5.5 &0.1 & 2.3 &0.1 \\
 ggF   &  23.8&2.6 & 75.7 &6.1 & 11.3 &2.2 & 13.2 &1.5 & 43.4&3.8 & 127.0&6.5 & 0.5 &0.2 & 0.3 &0.1 & 0.8 &0.3 \\
 \VH    &   0.2&0.2 & 6.0  &1.2 & 1.2  &0.9 & 0.7  &0.3 & 3.9 &0.8 & 28.9 &2.6 & \multicolumn{2}{r}{$<$0.1} & \multicolumn{2}{r}{$<$0.1} & \multicolumn{2}{r}{$<$0.1} \\
 \ttH   &   2.0&0.2 & 14.6 &0.7 & 0.3  &0.1 & 1.0  &0.1 & 5.7 &0.3 & 20.2 &0.5 & \multicolumn{2}{r}{$<$0.1} & \multicolumn{2}{r}{$<$0.1} & 0.4 &0.1 \\
\hline
\hline
\end{tabular}}
\end{table}

\section{Systematic uncertainties}
\label{sec:uncertainties}
The systematic uncertainties for the background and signal expectations are divided into experimental and theoretical uncertainties.
The uncertainties discussed below affect only the simulation-based signal and background predictions.
They do not affect the non-resonant background estimates because those estimates are derived from data.
All uncertainties are propagated to the BDT input variables and then to the final likelihood fits, with the exception of the luminosity uncertainty, which is taken as a constant uncertainty.
In the likelihood fits for signal extraction, the uncertainties affect the $m_{bb}$ spectrum modeling and normalization of signal processes in each region, as well as the $m_{bb}$ spectrum modeling of the \Zboson boson background.  The impact of the uncertainties on the BDT output and $m_{bb}$ shape are determined together and fully correlated. 
Uncertainties from sources common to the signal and \Zboson boson background are treated as correlated between the two.

\subsection{Experimental uncertainties}
\label{sec:expuncert}

The uncertainty in the integrated luminosity is 2.2\% for the \textit{all-hadronic} channels and 2.1\% for the \textit{photon} channel with the difference due to the small difference in luminosity between the channels.
It is derived, following a methodology similar to that detailed in Ref.~\cite{DAPR-2013-01} from a  calibration of the luminosity scale using $x$\nobreakdash--$y$ beam-separation scans performed in August 2015 and May 2016.
This systematic uncertainty is applied to all physics processes estimated with simulation samples.

The most prominent sources of jet-related uncertainty are the uncertainties in the jet energy scale (JES) and jet energy resolution (JER).
The JES uncertainty is determined primarily by using \Zboson, photon, and multijet \pT-balancing techniques in data~\cite{PERF-2016-04}.
The per-jet uncertainty in the energy scale varies from approximately 1\% to 5\% for the jets considered in this analysis.
The systematic uncertainties of the additional energy corrections specific to $b$-jets are found to be negligible.

The JER uncertainties are also determined \textit{in situ} via \Zboson, photon, and dijet \pT-balancing techniques~\cite{PERF-2016-04}. 
The systematic uncertainty due to the JER is calculated by increasing the resolution within its uncertainties, smearing the jet energy by the resulting change in resolution, and comparing the result to the nominal shape and normalization in simulation. 
The signal mass resolution varies by 3\% to 4\% due to the systematic uncertainty in the jet energy resolution.

The uncertainties related to the $b$-tagging of jets are implemented as variations of simulation correction factors (scale factors). These scale factors and their associated uncertainties are determined from data using \ttbar events,  $W+c$ and  $D^*$ events, and multijet data~\cite{PERF-2012-04,ATL-PHYS-PUB-2016-012}.
The systematic uncertainties for each jet are propagated to a total event uncertainty.
To simplify the computation and reduce the number of significant uncertainties, a principal-component analysis is performed over all of the contributing uncertainties to generate a reduced set of nuisance parameters.
For $b$-jets, the uncertainty is approximately 2\%, while it is 10\% for $c$-jets and 30\% for light jets.  Scale factors for the online $b$-tagging algorithms and their uncertainties are derived relative to the offline algorithms and applied to $b$-jets.  The uncertainties are typically 2--5\%.   

The uncertainty due to the jet vertex tagging requirement is measured in $\Zboson(\to \ell^+\ell^-)$ + 1-jet events.
The uncertainty per event is less than 2\%~\cite{ATLAS-CONF-2014-018}.  

To estimate the effects of uncertainties in the number of tracks associated with a jet, tracks are removed or added according to the tracking efficiency and fake-rate estimates~\cite{PERF-2015-08,}.
Uncertainties in the modeling of track multiplicity are derived from the measurement of the charged-particle multiplicity inside jets from $\sqrt{s} = \SI{8}{\TeV}$ $pp$ collisions~\cite{STDM-2015-12}.
These effects lead to an uncertainty in the average number of charged particles associated with the jet of approximately 10\%.

In the \textit{photon} channel, the analysis is not highly sensitive to photon energy uncertainties, so multiple sources of electromagnetic energy scale and resolution uncertainties are combined into a set of just two parameters. The uncertainties were derived from calibration studies in data and  data to simulation comparisons~\cite{ATL-PHYS-PUB-2016-015, ATL-PHYS-PUB-2016-014}.
A data-driven  correction is applied to account for a shift between the data and simulation distributions of the photon isolation energy.
The  difference between the uncorrected and corrected isolation energy is taken as a systematic uncertainty. 

Systematic uncertainties from electrons and muons are negligible and hence are neglected.

\subsection{Theoretical uncertainties}
\label{sec:theoryuncert}
The value of the $H\ra\bbbar$ branching ratio and its uncertainty are from the recommendations of the LHC Higgs Cross Section Working Group for $m_H = \SI{125}{\GeV}$~\cite{deFlorian:2016spz} and 
are calculated by the HDECAY program~\cite{Djouadi:1997yw}.
Uncertainties in the cross section and acceptance for VBF and ggF signals due to the missing higher-order terms in perturbative QCD calculations are evaluated by varying the choice of renormalization scale and factorization scale independently by factors of 0.5 and 2.0.  
Specific uncertainties are applied for ggF events with additional radiation which generates a VBF-like topology.  
The ggF events are classified as VBF-like if they have at least two additional jets with an invariant mass greater than \SI{400}{\GeV}.  The uncertainties in the cross section are approximately 20\% in this phase space.     
The total cross-section and acceptance uncertainties affect the signal yields by 4--15\% in the \textit{all-hadronic} channels and 10--16\% in the \textit{photon} channel.
Uncertainties in the cross section and acceptance due to the choice of PDF are evaluated by varying the error eigenvectors of the nominal PDFs.  
They result in 5--10\% uncertainties in the signal yields.

The uncertainty from the parton-shower and underlying-event models is estimated by comparing the nominal sample, which uses \PYTHIA 8.2 for parton showering, with an alternative sample using \HERWIG 7.0 for parton-shower generation. 
This uncertainty is 4--12\%.  These uncertainties are also propagated to the $m_{bb}$ shape.  

The contributions of the \VH and \ttH Higgs boson production modes to the \textit{all-hadronic} channels' signal regions are small in the most sensitive signal regions (0.2--3\%)
and rise to 20\% in the least sensitive regions.
The contribution from these processes is included in the total Higgs boson yield, and 100\% uncertainty is taken for their relative contribution.
In the \textit{photon} analysis the ggF and \ttH contributions are small, and 100\% uncertainty is assumed.
The yield from these processes is added to the Higgs boson yields from VBF and \VH processes.

\subsection{Non-resonant and \Zboson boson background uncertainties}
\label{sec:fituncert}

The uncertainty due to the non-resonant background modeling is included by determining the largest spurious signal induced in Asimov datasets derived with alternative functions which describe the data sidebands equally well.  
These alternative functions must pass the $\chi^2$ and $F$-test as described in Section~\ref{sec:backgrounds}.  
The size of the spurious signal is taken as the uncertainty and is typically 20--30\% of the expected Higgs boson signal.   This uncertainty is included in the total experimental uncertainties. 

The uncertainties due to the \Zboson boson background fall into two categories.  
Experimental uncertainties in the observed \Zboson boson resonance shape are determined as described in Section~\ref{sec:expuncert}.    
Normalization uncertainties are determined from the fit.

\section{Fits for Higgs boson production}
\label{sec:results}
The inclusive Higgs boson signal strength $\muH$ and the VBF-specific strength $\muVBF$ are extracted from an extended maximum-likelihood fit to the $b$-tagged dijet invariant mass spectrum $m_{bb}$ in data.
The \textit{two-central} and \textit{four-central} channels use a joint binned likelihood fit with a bin size of \SI{0.5}{\GeV}.  The signal strength is common to the two channels.  The \textit{photon} channel, which has  fewer events, uses an unbinned fit to maximize the sensitivity.  The fit range is $\SI{80}{\GeV} <  m_{bb}  < \SI{200}{\GeV}$ for the \textit{two-central} and \textit{four-central} channels and $\SI{50}{\GeV} <  m_{bb} < \SI{250}{\GeV}$ for the \textit{photon} channel.  The different lower mass bounds for the two fits is because the \textit{photon} channel has lower jet thresholds.  For the  \textit{two-central} and \textit{four-central} channel fits, there is no benefit to extending the fit range beyond an $m_{bb}$ of \SI{200}{\GeV}.

In all cases, the likelihood is built from the product of Poisson probability terms across all channels and BDT regions with three contributions: non-resonant background, \Zboson boson events,  and Higgs boson signal events.  The parameterization of these contributions is described in Section~\ref{sec:backgrounds}.  The likelihood includes terms for systematic uncertainties implemented as nuisance parameters.
The nuisance parameters describe the systematic uncertainties discussed in Section~\ref{sec:uncertainties} and are parameterized by Gaussian or log-normal priors.  
Each prior constrains a nuisance parameter to its nominal value within its associated uncertainty.

The strength of the Higgs boson signal, either inclusive or VBF-specific, is the parameter of interest. 
Other free parameters include the shape parameters of the non-resonant background and the normalizations of the non-resonant and \Zboson boson backgrounds in each region.
Signal-injection tests confirmed the linearity of the fit with no bias.
The \textit{all-hadronic} and \textit{photon} results are also combined in a simultaneous likelihood fit with the signal strength treated as correlated across all analysis regions.
In the case of the inclusive extraction of $\muH$, all production mechanisms (VBF, ggF, \VH, and \ttH) are considered as signal, and their ratios are fixed to the SM predictions.
In the case of the VBF-only extraction of $\muVBF$, all channels include the contributions of ggF, \ttH, and \VH as nuisance parameters constrained to their Standard Model expectations with the uncertainties described in Section~\ref{sec:theoryuncert}.

With the exception of the $b$-tagging uncertainties, the experimental uncertainties are treated as fully correlated between the channels.
The $b$-tagging uncertainties for both the offline and online algorithms are taken as fully uncorrelated between different working points.
Treating them as correlated or uncorrelated has no impact on the overall result or uncertainty.
Theoretical uncertainties, including those in the QCD scale of the VBF process, the parton showering, the PDFs, and the \ttH yield, are correlated.
In general, background systematic uncertainties such as non-resonant background normalization/parameterization, \Zboson boson normalization, and spurious signals are specific to each channel and consequently not correlated. 
Systematic uncertainties related to the fit procedure are characterized by  spurious-signal nuisance parameters.  
These are treated as uncorrelated across the signal regions. 

\begin{figure}[htbp]
  \centering
 \includegraphics[width=0.48\textwidth]{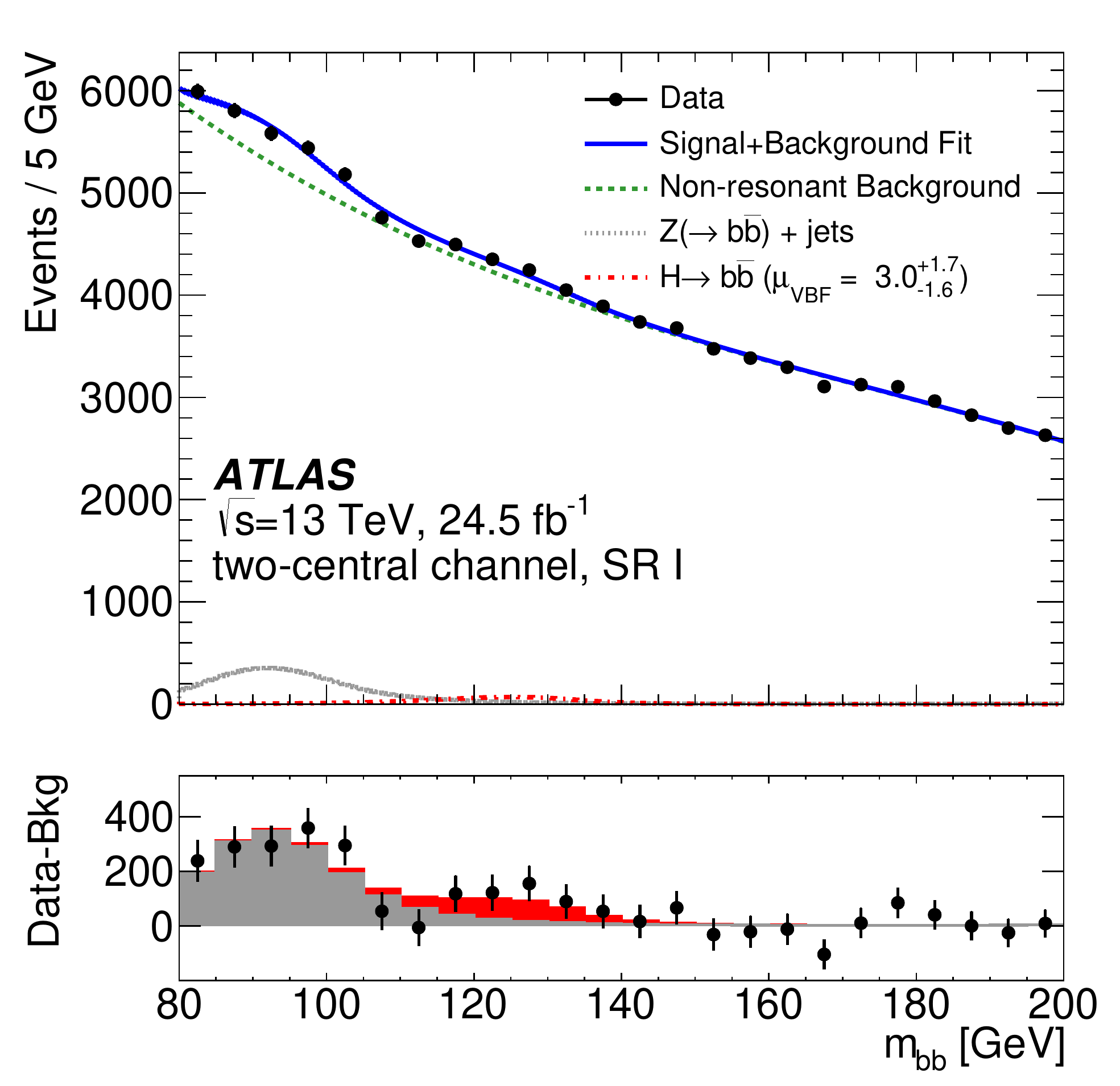}
 \includegraphics[width=0.48\textwidth]{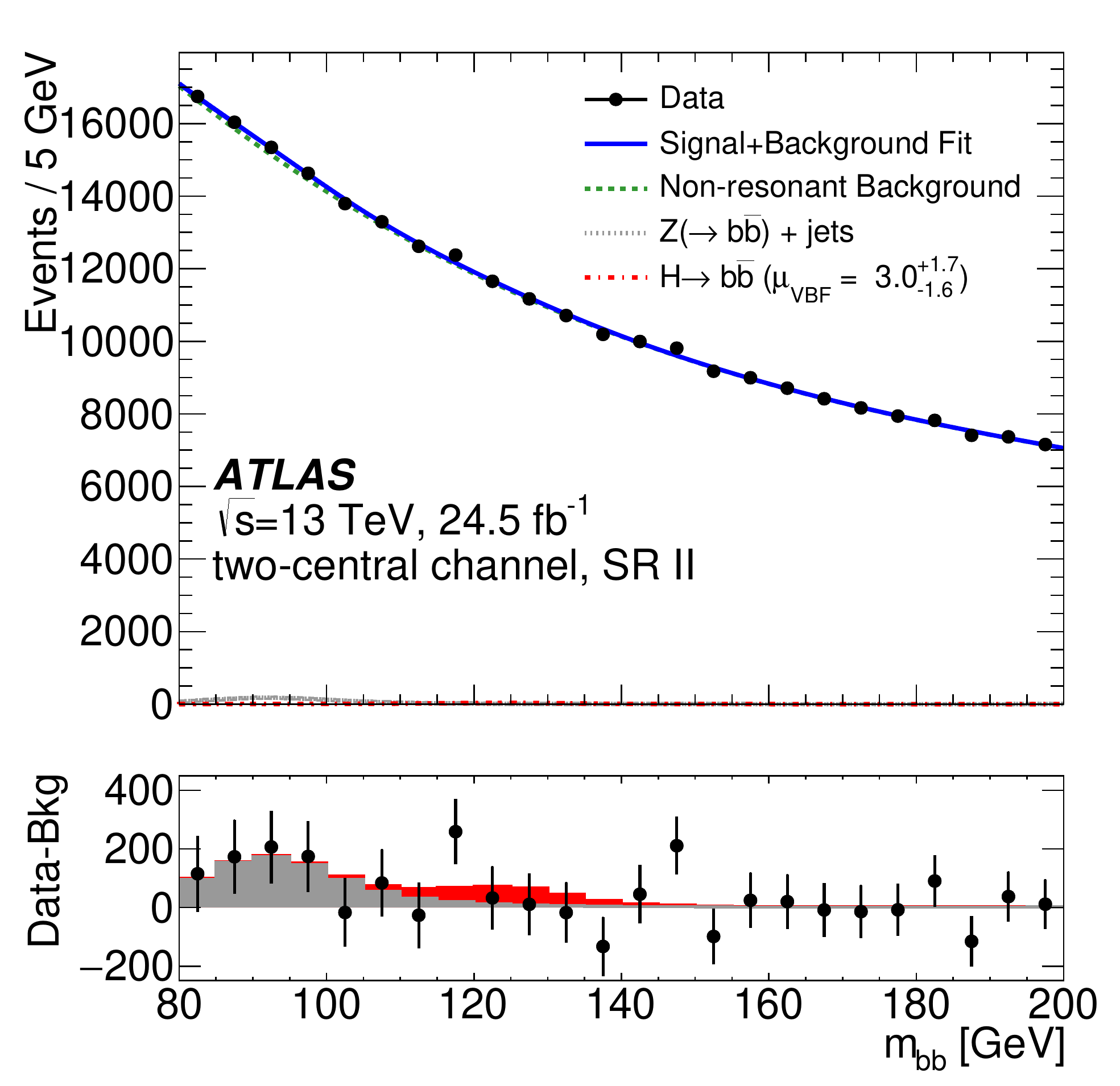}\\

\caption{Data and fit model comparison for the combined profile likelihood fit for \muVBF in the \textit{two-central} channel signal regions.  The combined fit includes all signal regions for the \textit{all-hadronic} and \textit{photon} channels. 
The fitted continuum background is shown with a dashed green line, the fitted $Z$ boson background with a dotted gray line, and the fitted Higgs boson signal with a dash-dotted red line. The total fit is displayed with a solid blue line. The bottom panels show the residuals of the data relative to the continuum background fit, along with the simulated $Z$ boson background and Higgs boson signal normalized to the fitted signal strengths.  Only statistical uncertainties are shown. }
  \label{fig:higgsfit_2cen}
\end{figure}

\begin{figure}[htbp]
  \centering
\includegraphics[width=0.48\textwidth]{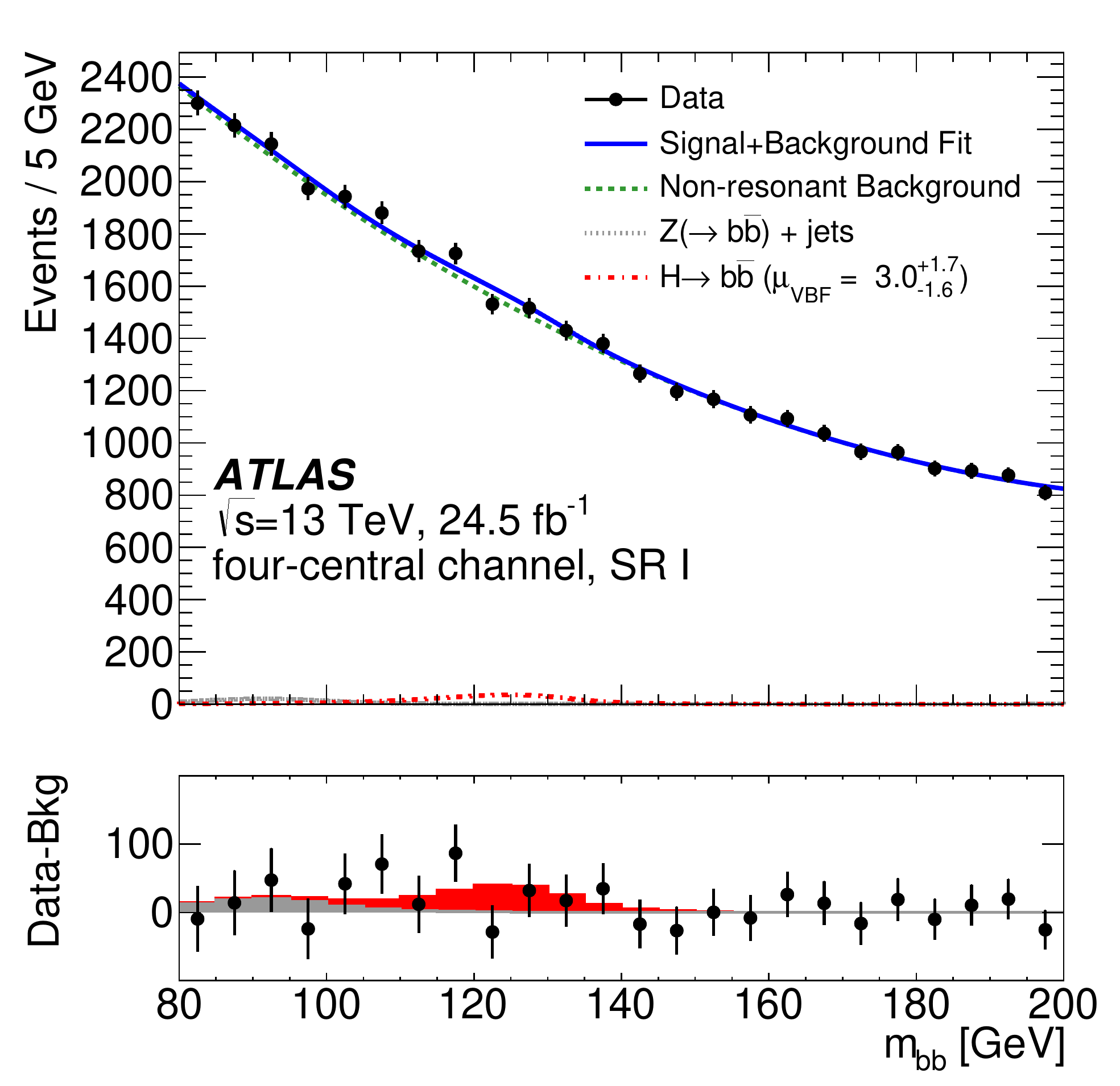}
 \includegraphics[width=0.48\textwidth]{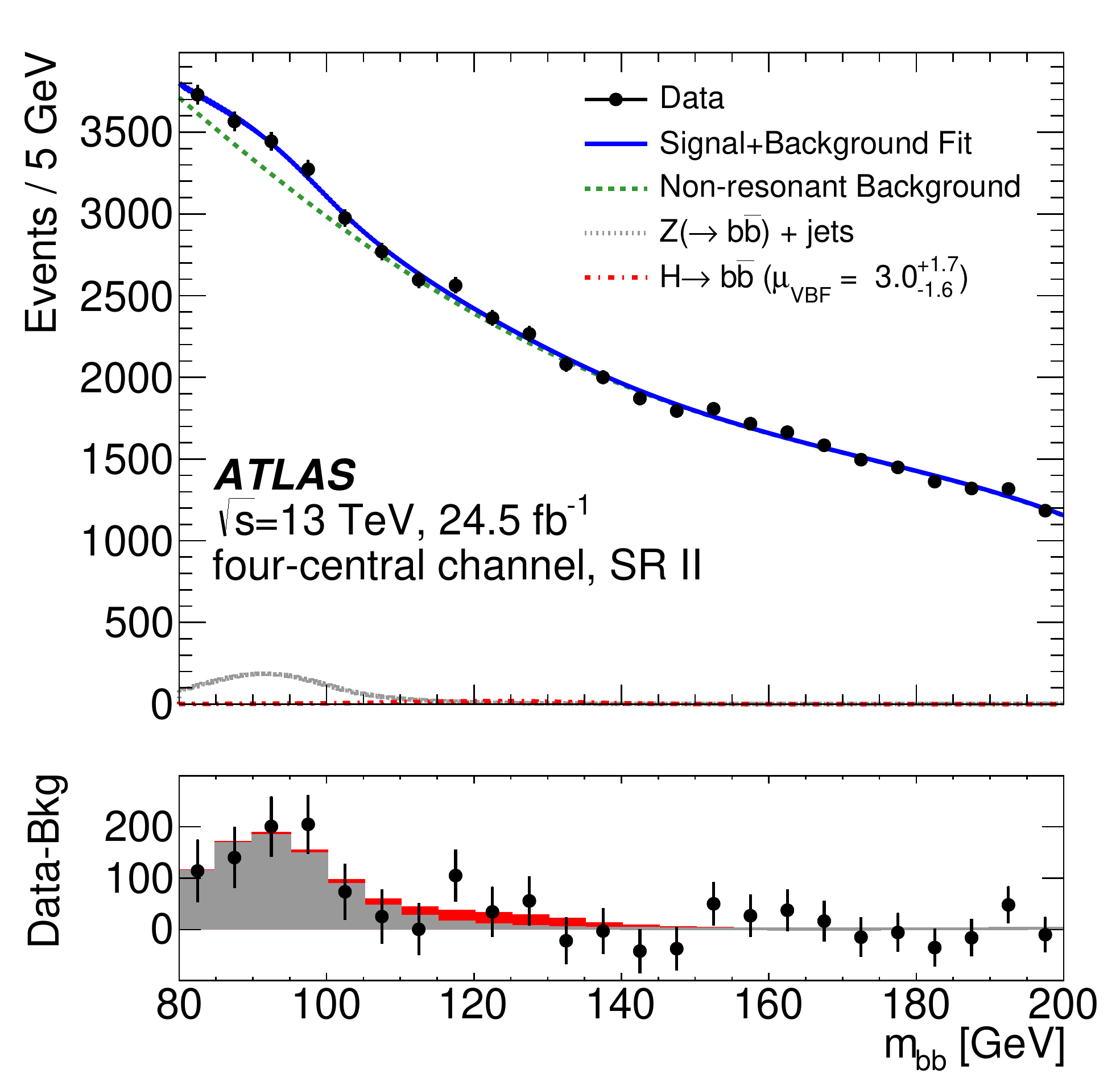}\\
 \includegraphics[width=0.48\textwidth]{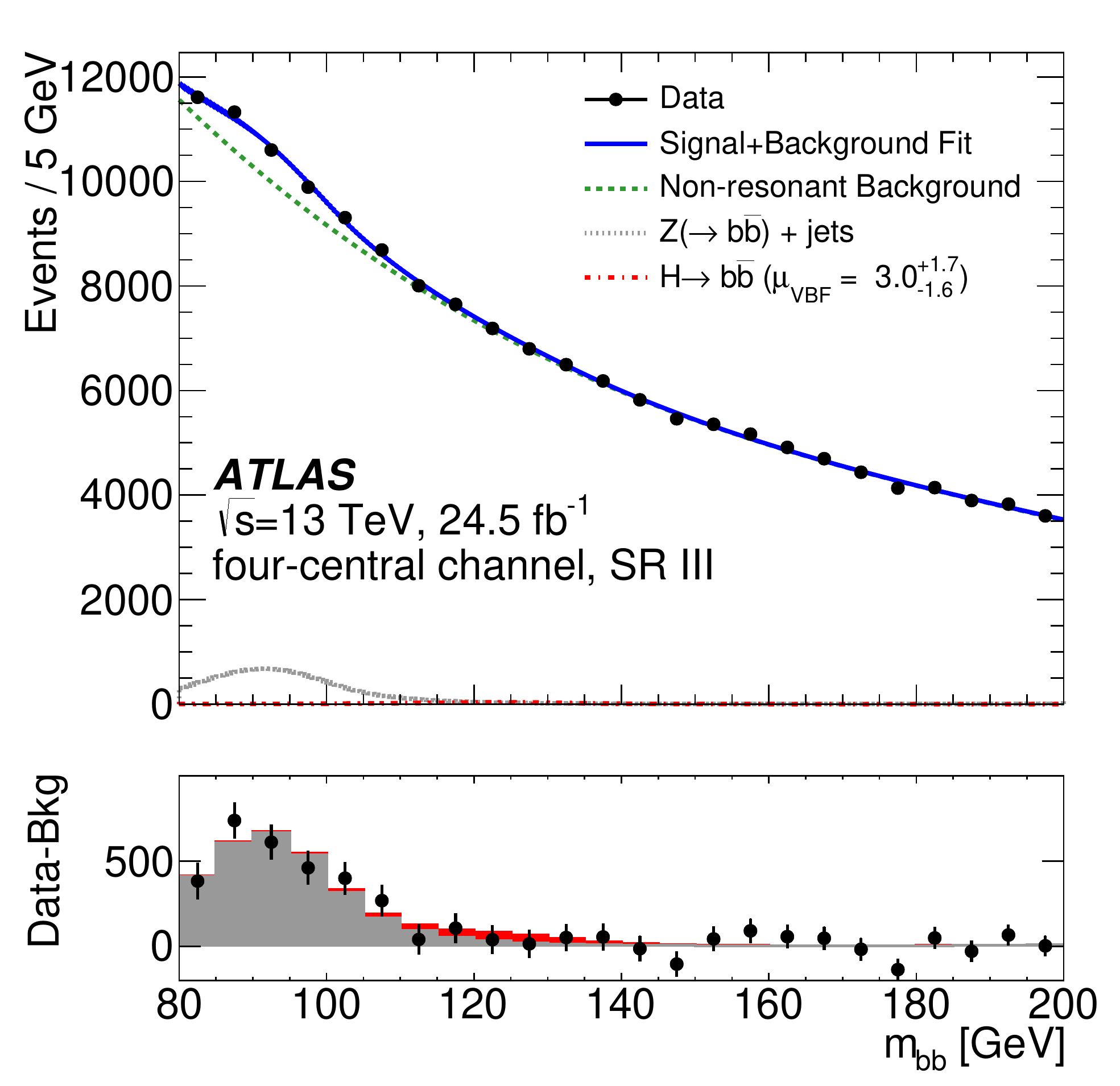}
 \includegraphics[width=0.48\textwidth]{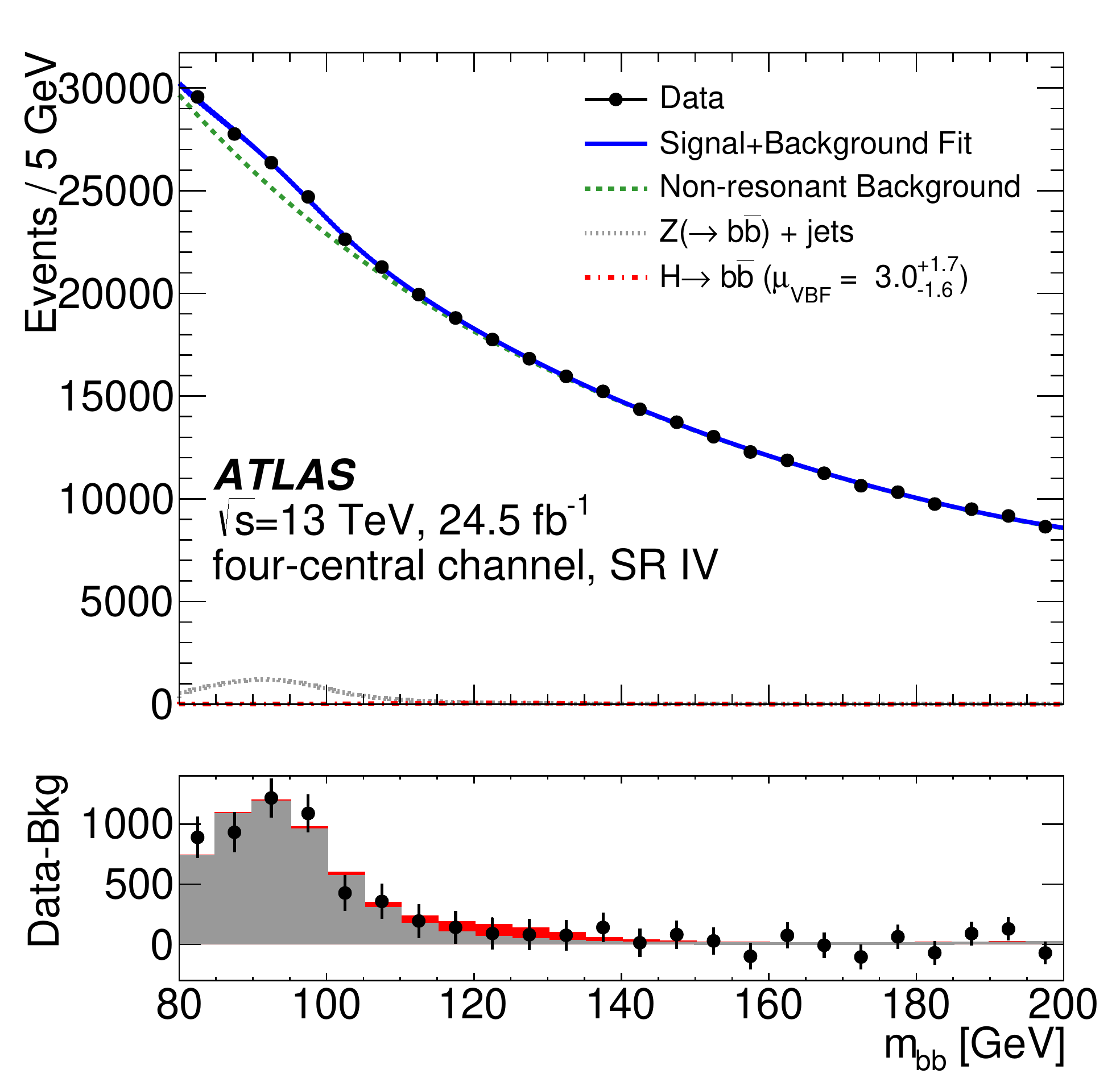}\\

\caption{Data and fit model comparison for the combined profile likelihood fit for \muVBF in the \textit{four-central} channel signal regions.  The combined fit includes all signal regions for the \textit{all-hadronic} and \textit{photon} channels.  The fitted continuum background is shown with a dashed green line, the fitted $Z$ boson background with a dotted gray line, and the fitted Higgs boson signal with a dash-dotted red line.  The total fit is displayed with a solid blue line. The bottom panels show the residuals of the data relative to the continuum background fit, along with the simulated $Z$ boson background and Higgs boson signal normalized to the the fitted signal strengths. Only statistical uncertainties are shown.}
  \label{fig:higgsfit_4cen}
\end{figure}

\begin{figure}[htbp]
\centering

\includegraphics[width=0.48\textwidth]{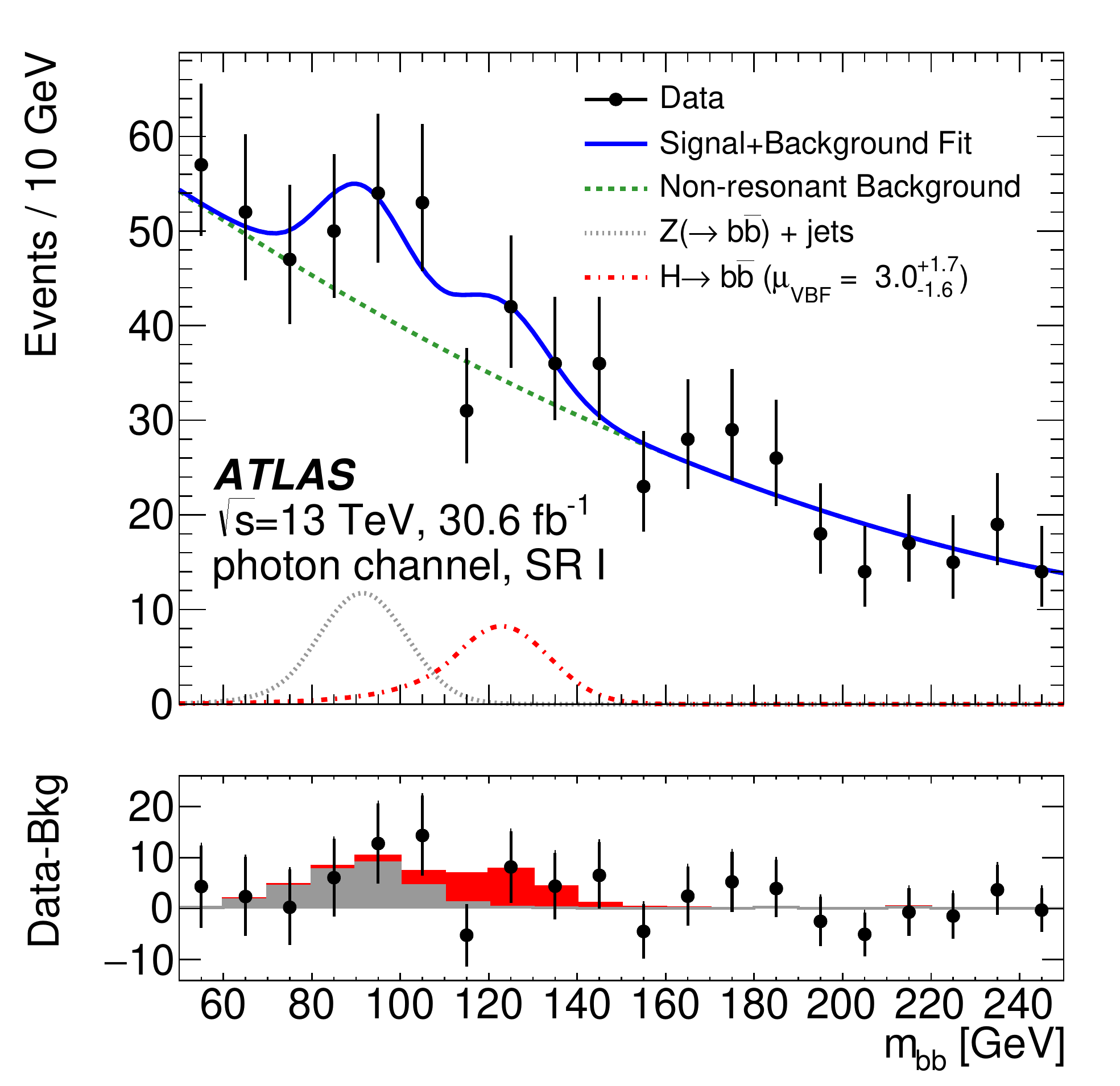}
\includegraphics[width=0.48\textwidth]{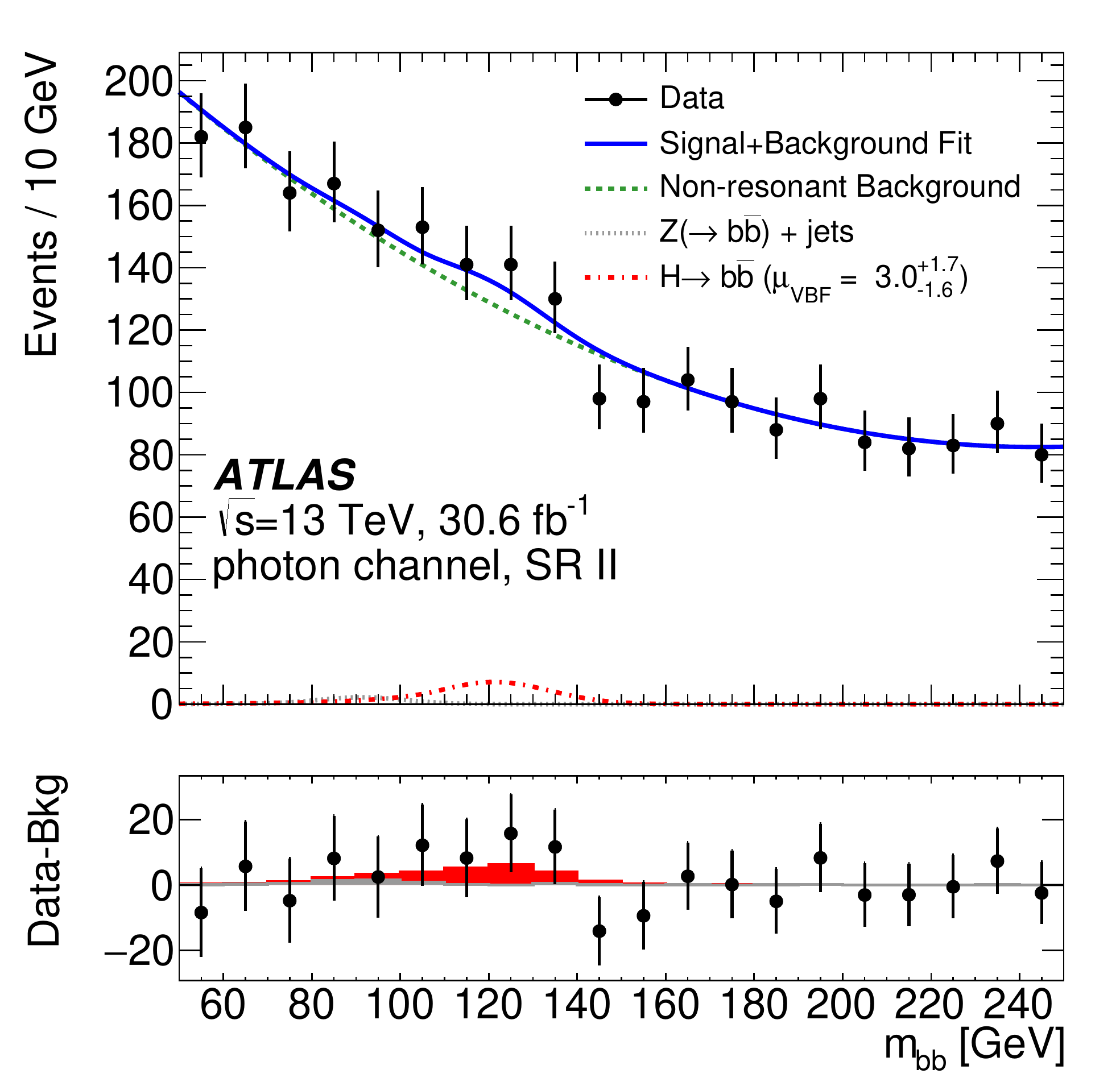}
\includegraphics[width=0.48\textwidth]{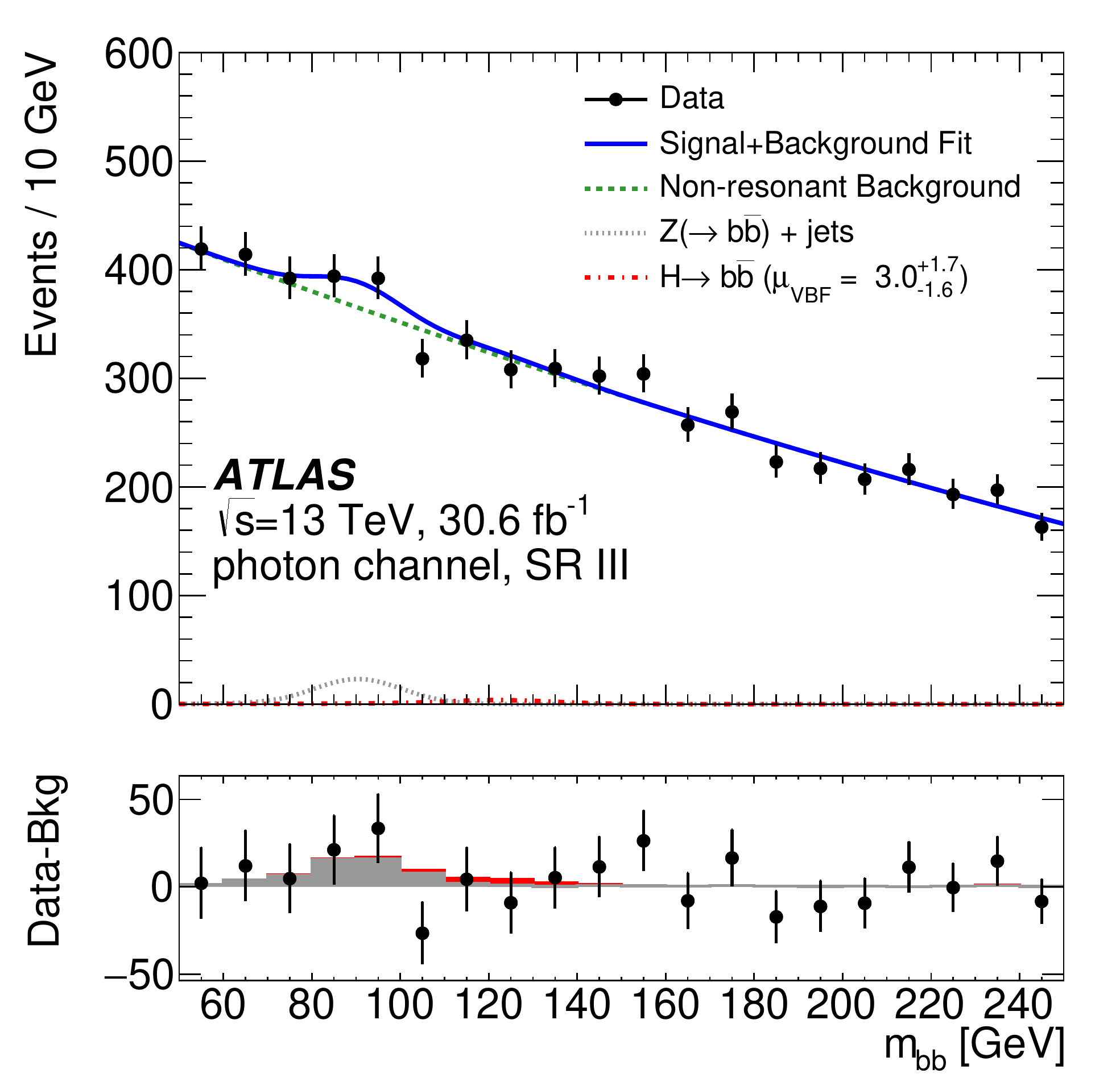}
\caption{Data and fit model comparison for the combined profile likelihood fit for \muVBF in the \textit{photon} channel signal regions.  The combined fit includes all signal regions for the \textit{all-hadronic} and \textit{photon} channels. The fitted continuum background is shown with a dashed green line, the fitted $Z$ boson background with a dotted gray line, and the fitted Higgs boson signal with a dash-dotted red line.  The total fit is displayed with a solid blue line. The bottom panels show the residuals of the data relative to the continuum background fit, along with the simulated $Z$ boson background and Higgs boson signal normalized to the fitted signal strengths. Only statistical uncertainties are shown.}
\label{fig:mbb_postfit_photon}
\end{figure}

The $m_{bb}$ invariant mass distributions after the combined $\muVBF$ fits are shown
in Figures~\ref{fig:higgsfit_2cen}--\ref{fig:mbb_postfit_photon} for each region and each channel. 
The Higgs boson signal, \Zboson boson background, and non-resonant background yields in the Higgs boson mass window of $\SI{100}{\GeV} < m_{bb} < \SI{140}{\GeV}$ after performing the combined fit are shown in Table~\ref{tab:postfit_yields_vbffit}.

A test statistic based on the profile likelihood function is used to determine the probability that the dataset is compatible with the Higgs boson signal hypothesis.
Distributions of the test statistic under the signal and null (background-only) hypotheses are estimated using asymptotic approximations~\cite{Cowan:2010js}.
As no statistically significant signal is observed, the $\text{CL}_{\text{s}}$ technique~\cite{Read:2002hq} is used to derive 95\% CL upper limits on $H\rightarrow b\bar{b}$ production in both the inclusive and VBF channels.
The likelihood fit results for the Higgs boson normalization are shown in Table~\ref{tab:higgs_significances_limits}, both for the individual channels and for the combined fits.
The results are consistent with Standard Model expectations within the uncertainties. 
A summary of the uncertainties is shown in Table~\ref{tab:systematic_uncertainty_summary}.  
The effect of data statistical uncertainty on the Higgs signal strength is derived by fixing all nuisance parameters to their best-fit values and taking the differences between the central value and the 1$\sigma$ interval for the measured Higgs signal strength. 
The effect from non-resonant background parameters is then derived as the difference in quadrature between the uncertainty effects on the Higgs signal strength derived by floating and fixing the corresponding nuisance parameters.
A similar procedure is performed to calculate the impact of the other uncertainties. 
The total uncertainty is dominated by statistical uncertainties, with important contributions from the determination of the non-resonant background parameters and $Z$ normalization due to the weak constraining power of the low $m_{bb}$ sideband.  The experimental systematic uncertainties, as defined in Section~\ref{sec:expuncert}, also contribute significantly to the total uncertainty.  
Of these, the leading uncertainties are due to the JES and JER uncertainties, followed by $b$-tagging uncertainties.
The spurious signal contributes an uncertainty of less than 0.1 for both the individual and combined signal extractions.

\begin{figure}[htbp]
  \centering
\includegraphics[width=0.49\textwidth]{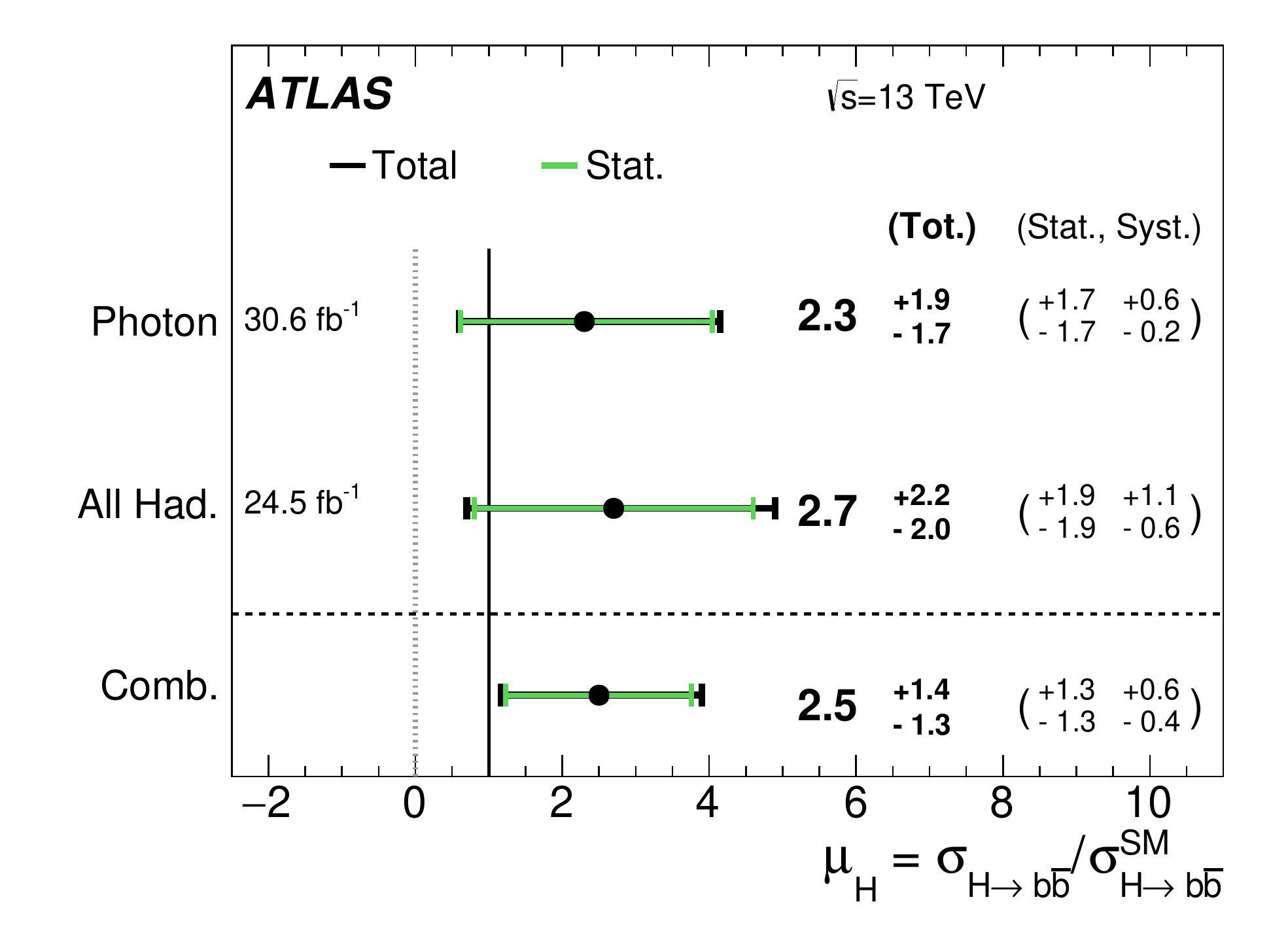}
\includegraphics[width=0.49\textwidth]{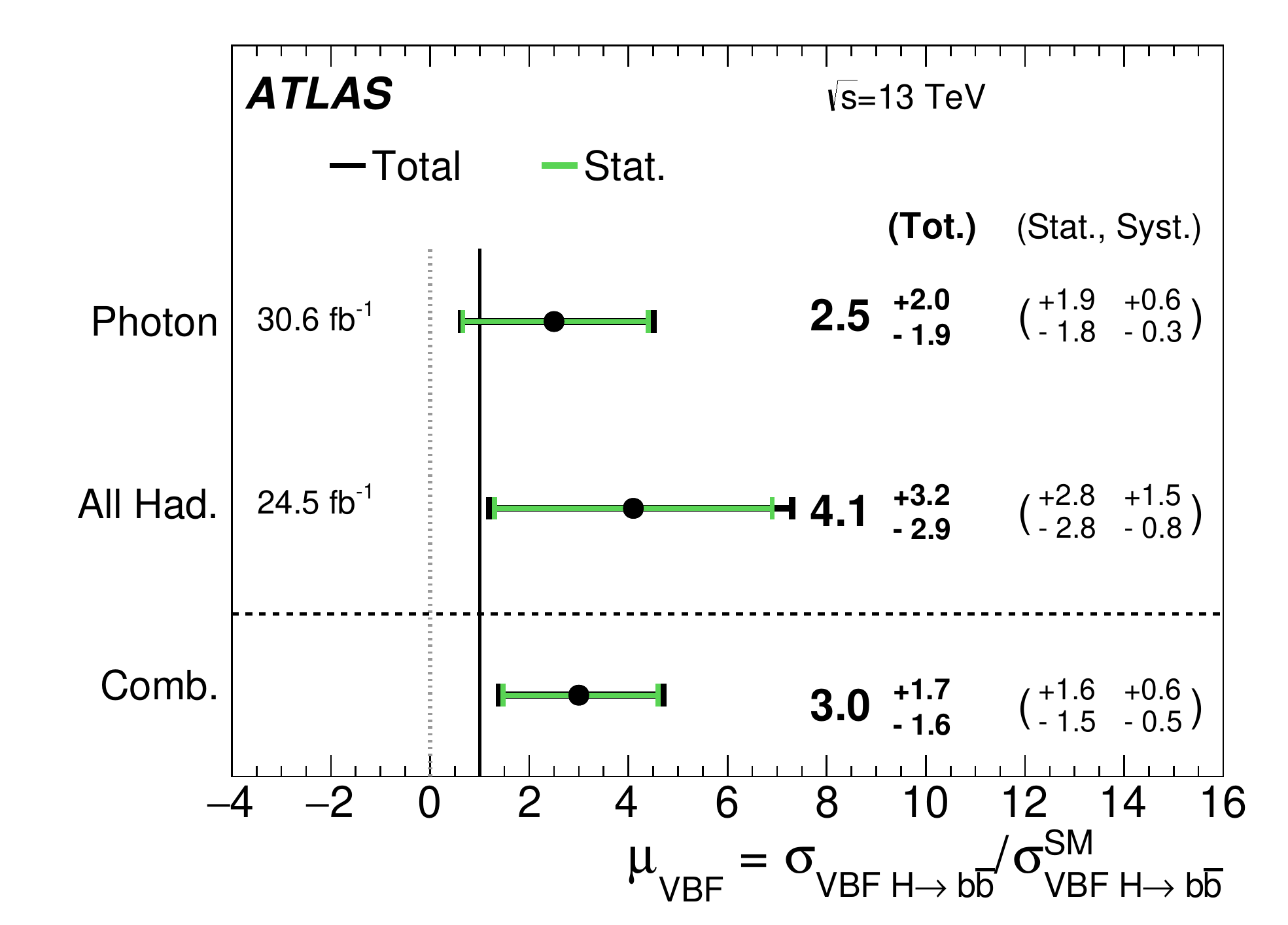}

\caption{Summary of the \textit{all-hadronic}, \textit{photon} and combined results for the fitted signal strength parameters $\muH$ (left) and $\muVBF$ (right).
  The inner error bars show the statistical uncertainties. The outer error bars show the total uncertainties.}
  \label{fig:summary}
\end{figure}

The results for the extraction of $\muH$ and $\muVBF$ are also shown in Table~\ref{tab:higgs_significances_limits} and displayed in Figure~\ref{fig:summary}. 
The observed significances and signal strengths are higher than the expected significances for all channels. 
The observed significances of both the inclusive and VBF-only production are 1.9$\sigma$, compared with 0.8$\sigma$ expected for the inclusive production and 0.7$\sigma$ expected for the VBF production. The observed signal strength, $\muH$, is 2.5$^{+1.4}_{-1.3}$ for inclusive production, as compared with  1.0$\pm1.2$ expected.  For VBF production,  $\muVBF$  is observed to be  3.0$^{+1.7}_{-1.6}$, which can be compared with an expectation of $1\pm1.5$. 

\begin{table}[htbp]
   \caption{Numbers of signal, background and data events within the Higgs boson mass window of $\SI{100}{\GeV} < m_{bb} < \SI{140}{\GeV}$. Signal and background yields are derived from the combined fit for the extraction of \muVBF. Uncertainties include both the statistical and systematic uncertainties.}
\label{tab:postfit_yields_vbffit}
\centering
\def\arraystretch{1.4}
\resizebox{\textwidth}{!}{
   \begin{tabular}{ l |r@{$^{+}_{-}$}l|r@{$^{+}_{-}$}l|r@{$^{+}_{-}$}l|r@{$^{+}_{-}$}l|r@{$^{+}_{-}$}l|r@{$^{+}_{-}$}l|r@{$^{+}_{-}$}l|r@{$^{+}_{-}$}l|r@{$^{+}_{-}$}l}
\hline
\hline
Channel      & \multicolumn{4}{c|}{\textit{Two-central}} & \multicolumn{8}{c|}{\textit{Four-central}} & \multicolumn{6}{c}{\textit{Photon}} \\ \hline
Region       & \multicolumn{2}{c|}{SR I}           & \multicolumn{2}{c|}{SR II  }         & \multicolumn{2}{c|}{SR I }  & \multicolumn{2}{c|}{SR II} & \multicolumn{2}{c|}{SR III} & \multicolumn{2}{c|}{SR IV}  & \multicolumn{2}{c|}{SR I}  & \multicolumn{2}{c|}{SR II}   & \multicolumn{2}{c}{SR III}   \\
\hline
Higgs boson & 340&$^{120}_{130}$ & 165&$^{50}_{29}$ & 167&$^{60}_{58}$ & 101&$^{40}_{21}$ & 183&$^{50}_{46}$ & 304&$^{100}_{51}$ & 21.1&$^{7.7}_{7.1}$ & 20.1&$^{9.5}_{7.2}$ & 10.6&$^{7.8}_{4.1}$ \\
$Z$+jets ($Z\gamma$) & 470&$^{140}_{180}$ & 230&$^{210}_{230}$ & 22&$^{80}_{22}$ & 197&$^{90}_{95}$ & 720&$^{190}_{180}$ & 1,260&$^{270}_{250}$ & 5.8&$^{3.3}_{3.6}$ & 1.1&$^{5.8}_{1.1}$ & 9.8&$^{7.8}_{7.9}$ \\
Non-resonant bkg & 34,620&$^{310}_{280}$ & 95,620&$^{420}_{420}$ & 12,870&$^{150}_{190}$ & 19,340&$^{200}_{240}$ & 59,340&$^{340}_{340}$ & 146,930&$^{630}_{510}$ & 140.4&$^{6.1}_{6.8}$ & 518&$^{10}_{13}$ & 1,296&$^{18}_{19}$ \\
 [2px] \hline
 Data & \multicolumn{1}{r}{35,496} && \multicolumn{1}{r}{95,802} && \multicolumn{1}{r}{13,139} && \multicolumn{1}{r}{19,611} && \multicolumn{1}{r}{60,314} && \multicolumn{1}{r}{148,413} && \multicolumn{1}{l}{162} && \multicolumn{1}{r}{565} && \multicolumn{1}{r}{1,270} \\
\hline
\hline
\end{tabular}}
\end{table}

\begin{table}[hbtp]
\small
\caption{Expected and observed results for the Higgs boson production rate, for both inclusive production and VBF production only, relative to the Standard Model prediction. Where the results are reported by channel, the fit is performed with that channel only. The limits shown refer to 95\% CL upper limits. }
\label{tab:higgs_significances_limits}

\centering
\def\arraystretch{1.4}
\begin{tabular}{l|rrr|rrr}
\hline
\hline
\multirow{2}{*}{Results} & \multicolumn{3}{c|}{Inclusive production} & \multicolumn{3}{c}{VBF production} \\
\cline{2-7}
& \textit{All-hadronic} & \textit{Photon} & Combined & \textit{All-hadronic} & \textit{Photon} & Combined\\
\hline
Expected significance & $0.5\sigma$ & $0.6\sigma$&  $0.8\sigma$ & $0.4\sigma$ & $0.6\sigma$&  $0.7\sigma$  \\
Observed significance & $1.4\sigma$ & $1.3\sigma$&  $1.9\sigma$ & $1.4\sigma$ & $1.4\sigma$&  $1.9\sigma$  \\
Expected limit on signal strength & $4.1^{+1.9}_{-1.2} $  & $3.4^{+1.5}_{-1.0}$ & $2.5^{+1.0}_{-0.7}$ & $5.9^{+2.6}_{-1.7} $  & $3.7^{+1.6}_{-1.0}$ & $3.0^{+1.3}_{-0.8}$ \\
Observed limit on signal strength & 6.8 & 5.5 &  4.8 & 9.7 & 6.1 & 5.9 \\
Expected signal strength  & $1.0\pm1.9$ & $1.0\pm1.7$ & $1.0\pm 1.2$ & $1.0\pm2.8$ & $1.0\pm1.8$ & $1.0\pm 1.5$   \\
Observed signal strength  & $2.7^{+2.2}_{-2.0}$ & $2.3^{+1.9}_{-1.7}$ & $2.5^{+1.4}_{-1.3}$ & $4.1^{+3.2}_{-2.9}$ & $2.5^{+2.0}_{-1.9}$ & $3.0^{+1.7}_{-1.6}$ \\
\hline
\hline

\end{tabular}
\end{table}

\begin{table}[hbtp]
\centering
\caption{Uncertainties and their effects on the Higgs boson signal strength in the combined fit
for both the inclusive production (\muH) and VBF-only production (\muVBF).
The combined fit includes all signal regions for the \textit{all-hadronic} channels and \textit{photon} channel. 
Uncertainties are grouped into statistical and systematic uncertainties.
}
\label{tab:systematic_uncertainty_summary}
\begin{tabular}{l|c|c}
\hline
\hline
Uncertainty    &$\sigma(\muH)$ &$\sigma(\muVBF)$ \\
\hline
Total stat. uncertainty             &  $+$1.3 $-$1.3  &   $+$1.6 $-$1.5  \\
\quad Data stat. uncertainty        &  $+$0.6 $-$0.6  &   $+$0.9 $-$0.9  \\
\quad Non-resonant bkg              &  $+$1.0 $-$1.0  &   $+$1.2 $-$1.2  \\
\quad $Z$+jets normalization          &  $+$0.5 $-$0.5  &   $+$0.5 $-$0.5  \\
\hline
Total syst. uncertainty             &  $+$0.6 $-$0.4  &   $+$0.6 $-$0.5  \\
\quad Higgs boson modeling     &  $+$0.3 $-$0.1  &   $+$0.2 $-$0.1  \\
\quad JES/JER                       &  $+$0.3 $-$0.2  &   $+$0.4 $-$0.2  \\
\quad $b$-tagging (incl. trigger)   &  $+$0.2 $-$0.1  &   $+$0.2 $-$0.1  \\
\quad Other exp. uncertainty        &  $+$0.4 $-$0.3  &   $+$0.4 $-$0.4  \\
\hline
Total                               &  $+$1.4 $-$1.3  &   $+$1.7 $-$1.6  \\
\hline \hline
\end{tabular}
\end{table}

\FloatBarrier

\section{Conclusions}
\label{sec:conclusion}
A search is presented for the Standard Model Higgs bosons produced through vector-boson fusion and decaying into $\bbbar$ using three distinct event signatures including production with an associated photon. The results use up to \SI{30.6}{\ifb} of LHC $pp$ data at $\sqrt{s} = \SI{13}{\TeV}$ collected with the ATLAS detector.

The combined observed (expected) 95\% CL upper limits on the Higgs boson production cross section times branching ratio are $4.8$ ($2.5^{+1.0}_{-0.7}$) times the Standard Model expectation for inclusive production, and $5.9$ ($3.0^{+1.3}_{-0.8}$) times the Standard Model expectation for VBF production.

The measured Higgs boson signal strength relative to the Standard Model prediction \muH for the three channels combined is $2.5^{+1.4}_{-1.3}$ . The combined VBF-only signal strength  \muVBF  is $3.0^{+1.7}_{-1.6}$.

\section*{Acknowledgments}


We thank CERN for the very successful operation of the LHC, as well as the
support staff from our institutions without whom ATLAS could not be
operated efficiently.

We acknowledge the support of ANPCyT, Argentina; YerPhI, Armenia; ARC, Australia; BMWFW and FWF, Austria; ANAS, Azerbaijan; SSTC, Belarus; CNPq and FAPESP, Brazil; NSERC, NRC and CFI, Canada; CERN; CONICYT, Chile; CAS, MOST and NSFC, China; COLCIENCIAS, Colombia; MSMT CR, MPO CR and VSC CR, Czech Republic; DNRF and DNSRC, Denmark; IN2P3-CNRS, CEA-DRF/IRFU, France; SRNSFG, Georgia; BMBF, HGF, and MPG, Germany; GSRT, Greece; RGC, Hong Kong SAR, China; ISF, I-CORE and Benoziyo Center, Israel; INFN, Italy; MEXT and JSPS, Japan; CNRST, Morocco; NWO, Netherlands; RCN, Norway; MNiSW and NCN, Poland; FCT, Portugal; MNE/IFA, Romania; MES of Russia and NRC KI, Russian Federation; JINR; MESTD, Serbia; MSSR, Slovakia; ARRS and MIZ\v{S}, Slovenia; DST/NRF, South Africa; MINECO, Spain; SRC and Wallenberg Foundation, Sweden; SERI, SNSF and Cantons of Bern and Geneva, Switzerland; MOST, Taiwan; TAEK, Turkey; STFC, United Kingdom; DOE and NSF, United States of America. In addition, individual groups and members have received support from BCKDF, the Canada Council, CANARIE, CRC, Compute Canada, FQRNT, and the Ontario Innovation Trust, Canada; EPLANET, ERC, ERDF, FP7, Horizon 2020 and Marie Sk{\l}odowska-Curie Actions, European Union; Investissements d'Avenir Labex and Idex, ANR, R{\'e}gion Auvergne and Fondation Partager le Savoir, France; DFG and AvH Foundation, Germany; Herakleitos, Thales and Aristeia programmes co-financed by EU-ESF and the Greek NSRF; BSF, GIF and Minerva, Israel; BRF, Norway; CERCA Programme Generalitat de Catalunya, Generalitat Valenciana, Spain; the Royal Society and Leverhulme Trust, United Kingdom.

The crucial computing support from all WLCG partners is acknowledged gratefully, in particular from CERN, the ATLAS Tier-1 facilities at TRIUMF (Canada), NDGF (Denmark, Norway, Sweden), CC-IN2P3 (France), KIT/GridKA (Germany), INFN-CNAF (Italy), NL-T1 (Netherlands), PIC (Spain), ASGC (Taiwan), RAL (UK) and BNL (USA), the Tier-2 facilities worldwide and large non-WLCG resource providers. Major contributors of computing resources are listed in Ref.~\cite{ATL-GEN-PUB-2016-002}.

\clearpage

\printbibliography



\clearpage

 
\begin{flushleft}
{\Large The ATLAS Collaboration}

\bigskip

M.~Aaboud$^\textrm{\scriptsize 34d}$,    
G.~Aad$^\textrm{\scriptsize 99}$,    
B.~Abbott$^\textrm{\scriptsize 124}$,    
O.~Abdinov$^\textrm{\scriptsize 13,*}$,    
B.~Abeloos$^\textrm{\scriptsize 128}$,    
D.K.~Abhayasinghe$^\textrm{\scriptsize 91}$,    
S.H.~Abidi$^\textrm{\scriptsize 164}$,    
O.S.~AbouZeid$^\textrm{\scriptsize 39}$,    
N.L.~Abraham$^\textrm{\scriptsize 153}$,    
H.~Abramowicz$^\textrm{\scriptsize 158}$
% ...
\bigskip
\\

% ...
$^{13}$Institute of Physics, Azerbaijan Academy of Sciences, Baku; Azerbaijan.\\
% ...
$^{34}$$^{(a)}$Facult\'e des Sciences Ain Chock, R\'eseau Universitaire de Physique des Hautes Energies - Universit\'e Hassan II, Casablanca;$^{(b)}$Centre National de l'Energie des Sciences Techniques Nucleaires (CNESTEN), Rabat;$^{(c)}$Facult\'e des Sciences Semlalia, Universit\'e Cadi Ayyad, LPHEA-Marrakech;$^{(d)}$Facult\'e des Sciences, Universit\'e Mohamed Premier and LPTPM, Oujda;$^{(e)}$Facult\'e des sciences, Universit\'e Mohammed V, Rabat; Morocco.\\
% ...
$^{39}$Niels Bohr Institute, University of Copenhagen, Copenhagen; Denmark.\\
% ...
$^{91}$Department of Physics, Royal Holloway University of London, Egham; United Kingdom.\\
% ...
$^{99}$CPPM, Aix-Marseille Universit\'e, CNRS/IN2P3, Marseille; France.\\
% ...
$^{124}$Homer L. Dodge Department of Physics and Astronomy, University of Oklahoma, Norman OK; United States of America.\\
% ...
$^{128}$LAL, Universit\'e Paris-Sud, CNRS/IN2P3, Universit\'e Paris-Saclay, Orsay; France.\\
% ...
$^{153}$Department of Physics and Astronomy, University of Sussex, Brighton; United Kingdom.\\
% ...
$^{158}$Raymond and Beverly Sackler School of Physics and Astronomy, Tel Aviv University, Tel Aviv; Israel.\\
% ...
$^{164}$Department of Physics, University of Toronto, Toronto ON; Canada.\\
% ...
$^{a}$ Also at Borough of Manhattan Community College, City University of New York, NY; United States of America.\\
$^{b}$ Also at Centre for High Performance Computing, CSIR Campus, Rosebank, Cape Town; South Africa.\\
$^{c}$ Also at CERN, Geneva; Switzerland.\\
$^{d}$ Also at CPPM, Aix-Marseille Universit\'e, CNRS/IN2P3, Marseille; France.\\
$^{e}$ Also at D\'epartement de Physique Nucl\'eaire et Corpusculaire, Universit\'e de Gen\`eve, Gen\`eve; Switzerland.\\
% ...
$^{*}$ Deceased

\end{flushleft}


\end{document}